\documentclass[lettersize,journal]{IEEEtran}
\usepackage{amsmath,amsfonts}
\usepackage{algorithmic}
\usepackage{algorithm}
\usepackage{array}
\usepackage[caption=false,font=normalsize,labelfont=sf,textfont=sf]{subfig}
\usepackage{textcomp}
\usepackage{stfloats}
\usepackage{url}
\usepackage{verbatim}
\usepackage{graphicx}
\usepackage{ragged2e}
\usepackage{longtable}
\usepackage{multirow}
\usepackage{threeparttable}
\usepackage{booktabs}
\usepackage{supertabular}
\usepackage{cite}
\usepackage{booktabs,tabularx,threeparttablex}  
\usepackage{booktabs,tabularx,threeparttable,array}

\usepackage{xcolor,colortbl}

\definecolor{g0}{gray}{0.92}
\definecolor{g1}{gray}{0.75}
\definecolor{g2}{gray}{0.55}

\usepackage{xcolor}
\usepackage[normalem]{ulem} 
\usepackage{hyperref}

\begin{document}

\title{Security and Privacy in O-RAN for 6G: A Comprehensive Review of Threats and Mitigation Approaches}

\author{Lujia Liang, Lei Zhang \\

\thanks{
Lujia Liang, Lei Zhang (corresponding author) are with the James Watt School of Engineering, University of Glasgow, UK. Email: l.liang.1@research.gla.ac.uk, Lei.Zhang@glasgow.ac.uk. 
}
}



\maketitle

\newpage

\begin{abstract}
Open Radio Access Network (O-RAN) is a major advancement in the telecommunications field, providing standardized interfaces that promote interoperability between different vendors' technologies, thereby enhancing network flexibility and reducing operational expenses. By leveraging cutting-edge developments in network virtualization and artificial intelligence, O-RAN enhances operational efficiency and stimulates innovation within an open ecosystem. In the context of 6G, the potential capabilities of O-RAN have been significantly expanded, enabling ultra-reliable low-latency communication, terabit-level data rates, and seamless integration of terrestrial and non-terrestrial networks. Despite these benefits, its open architecture paradigm also brings critical security and privacy challenges, which, if not addressed, could compromise network integrity and data confidentiality. This paper conducts a comprehensive investigation into the security vulnerabilities and privacy issues associated with the O-RAN architecture in the context of the evolving 6G landscape, systematically categorizing fundamental vulnerabilities, meticulously examining potential attack vectors, and assessing current and future threats. In addition, this study also examines the existing and emerging security mechanisms of O-RAN and reviews the ongoing standardization activities aimed at strengthening the O-RAN security framework.
\end{abstract}


\begin{table*}[t]
\centering
\caption{Selected key acronyms used in this survey.}
\label{tab:acronyms}
\small
\setlength{\tabcolsep}{3pt}
\renewcommand{\arraystretch}{1.1}

\begin{tabularx}{\textwidth}{@{} l >{\raggedright\arraybackslash}X c l >{\raggedright\arraybackslash}X @{}}
\toprule
\textbf{Acronym} & \textbf{Full name} & & \textbf{Acronym} & \textbf{Full name} \\
\midrule

O-RAN & Open Radio Access Network &&
OAuth 2.0 & OAuth 2.0 Authorization Framework \\

RAN & Radio Access Network &&
RBAC & Role-Based Access Control \\

O-CU & O-RAN Central Unit &&
MACsec & Media Access Control Security \\

O-DU & O-RAN Distributed Unit &&
NDS/IP & Network Domain Security for IP \\

O-RU & O-RAN Radio Unit &&
UE & User Equipment \\

BBU & Baseband Unit &&
gNB & next Generation NodeB \\

RIC & RAN Intelligent Controller &&
AMF & Access and Mobility Management Function \\

Non-RT RIC & Non-Real-Time RAN Intelligent Controller &&
SMF & Session Management Function \\

Near-RT RIC & Near-Real-Time RAN Intelligent Controller &&
PEP & Policy Execution Point \\

xApp & (O-RAN) xApp &&
KPI & Key Performance Indicator \\

rApp & (O-RAN) rApp &&
KPM & Key Performance Measurement \\

SMO & Service Management and Orchestration &&
URLLC & Ultra-Reliable Low-Latency Comms \\

O-Cloud & O-RAN Cloud &&
HRLLC & High-Reliability Low-Latency Comms \\

FCAPS & Fault, Config, Accounting, Perf, Security &&
eMBB & enhanced Mobile Broadband \\

M-plane & Management plane &&
mIoT & massive Internet of Things \\

NFV & Network Functions Virtualization &&
SAGIN & Space-Air-Ground Integrated Network \\

VNF & Virtual Network Function &&
NTN & Non-Terrestrial Network \\

CNF & Cloud-Native Network Function &&
LEO & Low Earth Orbit \\

SDN & Software Defined Networking &&
GEO & Geostationary Earth Orbit \\

COTS & Commercial off-the-shelf &&
HAPS & High-Altitude Platform System \\

CI/CD & Continuous Integration/Continuous Deployment &&
UAV & Unmanned Aerial Vehicle \\

DevSecOps & Dev, Security, and Operations &&
THz & Terahertz \\

SBOM & Software Bill of Materials &&
RIS & Reconfigurable Intelligent Surface \\

CVE & Common Vulnerabilities and Exposures &&
ISAC & Integrated Sensing and Communication \\

SSDF & Secure Software Development Framework &&
FL & Federated Learning \\

STRIDE & Spoofing, Tampering, Repudiation, Info Disclosure, Denial of Service, Elevation of Privilege &&
DRL & Deep Reinforcement Learning \\

DoS & Denial of Service &&
V2X & Vehicle-to-Everything \\

DDoS & Distributed Denial of Service &&
ITU & International Telecommunication Union \\

PKI & Public Key Infrastructure &&
ITU-R & ITU Radiocommunication Sector \\

TLS & Transport Layer Security &&
IMT-2020 & Int. Mobile Telecommunications 2020 \\

mTLS & Mutual Transport Layer Security &&
IMT-2030 & Int. Mobile Telecommunications 2030 \\

IPsec & Internet Protocol Security &&
3GPP & 3rd Generation Partnership Project \\

ETSI & European Telecom. Standards Institute &&
SDO & Standards Development Organization \\

nGRG & Next Generation Research Group &&
PDP & Policy Decision Point \\
& \\

\bottomrule
\end{tabularx}
\end{table*}

\section{Introduction}
\IEEEPARstart{M}{obile} communication networks are fundamental to global digital transformation and economic growth. The rapid proliferation of smart devices and mobile services has heightened demands for high-speed, reliable, and extensive network coverage. Applications like autonomous driving \cite{guo2022vehicular}, Industry 5.0 \cite{Maddikunta2021Industry}, intelligent healthcare \cite{Topol2019High-performance}, and smart cities \cite{gohar2021role, allam2021future} require ultra-low latency, high reliability, and robust data processing. To address these challenges, 6G aims to integrate AI-driven self-optimisation with advanced network slicing technology, supporting ultra-high speeds (typically ranging from 50~Gbps to 200~Gbps), wireless network latency of 0.1--1~milliseconds(ms), and massive connectivity (from $10^6$ to $10^8$ devices per square kilometre) \cite{ITU-R_M.2160-0}.

\subsection{Survey Background}

Radio Access Network (RAN) is a key component of mobile communication networks, primarily responsible for managing end-user data and voice connectivity. With the rapid evolution of communication technologies, the design and functionality of RANs have evolved to accommodate increasingly complex network architectures and diverse service requirements. In the early stages of mobile networks, RAN systems relied heavily on these fixed base stations, which were strategically located to cover specific geographic areas. However, as mobile networks transitioned from 2G to 3G, the demand for data services skyrocketed, driving advances in the RAN to accommodate higher data rates and wider coverage. During this period, base stations began to support more complex data transmission tasks, incorporating advanced signal processing and spectrum management techniques to improve communication quality and service efficiency. The arrival of the 4G era marked a major breakthrough in RAN development, especially in network resource management and signal optimization. The implementation of efficient spectrum utilization strategies such as all-IP transmission technology \cite{Chiussi2002Mobility} and orthogonal frequency division multiplexing (OFDM) \cite{Oltean2004An} has significantly increased network capacity and speed. In addition, RAN architectures have become more dynamic, supporting adaptive network configuration and resource allocation to effectively manage fluctuating network traffic and changing user behavior.

Nevertheless, the limited availability of wireless spectrum resources remains an insurmountable challenge, requiring efficient and flexible management of RANs to support the growing demand for mobile services \cite{kassem2015future}. In conventional mobile networks (4G and some non-cloudified 5G networks), the management of the RAN is typically performed by a centralized controller. And the RAN functions are primarily concentrated in base station equipment, particularly in the Baseband Unit (BBU), which is responsible for baseband signal processing and most of the protocol stack functions. This equipment is serviced by proprietary vendors, which limits network flexibility and scalability \cite{parvez2018survey}. In the 5G era, network architecture is beginning to develop towards a more decentralized and flexible direction. For example, by introducing the concept of virtual RAN (V-RAN) and the widespread application of cloud RAN (C-RAN) proposed in the 4G era \cite{checko2014cloud}, it is possible to dynamically allocate and optimize network resources in a software-based manner. In addition, in order to further enhance the scalability and flexibility of the network, the concept of Open Radio Access Network (O-RAN) has emerged. As shown in Table \ref{tab:ran_comparison}, O-RAN aims to achieve interoperability of RAN devices from different vendors through open and standardized interfaces, thus breaking the traditional RAN vendor lock-in and providing more choices and flexibility for mobile network operators (MNOs) \cite{Niknam2022IntelligentNetworks}. The implementation of O-RAN not only promotes network innovation and diversity, but also helps to reduce network deployment and operating costs, promoting a healthier and more competitive market environment. As mobile communication technologies continue to advance, future networks will require smarter and more flexible management mechanisms to cope with increasing service demands and complex network environments. O-RAN and related technological innovations will play a key role in realizing this goal, providing a solid network foundation for the global digital transformation. According to 5G Americas \cite{5gAmericasOpenRAN2023}, Open RAN is the fastest-growing segment of the RAN market, and Open RAN is transitioning from early greenfield deployments to broader deployments in existing networks. This transition is expected to accelerate beyond 2025, culminating in the deployment of approximately 1.3 million Open RAN base stations by the end of this decade.

\begin{table*}[t]
    \caption{Comparison of RAN Architectures}
    \label{tab:ran_comparison}
    \centering

    \begin{tabularx}{\textwidth}{@{} >{\bfseries\RaggedRight\arraybackslash}p{0.15\textwidth} 
                                      >{\RaggedRight\arraybackslash}X 
                                      >{\RaggedRight\arraybackslash}X 
                                      >{\RaggedRight\arraybackslash}X 
                                      >{\RaggedRight\arraybackslash}X @{}}
        \toprule
        Comparison Dimension & \textbf{Traditional RAN} & \textbf{C-RAN} & \textbf{vRAN} & \textbf{O-RAN} \\
        \midrule
        
        Vendor Dependency & 
        Single-vendor, tightly coupled hardware and software, difficult to replace. & 
        Still relies on a single vendor, centralized architecture reduces site dependency. & 
        Uses general-purpose hardware, but software remains vendor-controlled. & 
        Multi-vendor ecosystem, hardware-software decoupling, avoids vendor lock-in. \\
        \addlinespace
        
        Compatibility \& Integration & 
        Proprietary interfaces, difficult to integrate with other vendors. & 
        Still uses proprietary interfaces, but centralized processing improves resource sharing. & 
        Runs on general-purpose servers, but cross-vendor compatibility remains limited. & 
        Open interfaces allow interoperability among different vendors. \\
        \addlinespace
        
        Hardware \& Software Flexibility & 
        Hardware tightly coupled with software, difficult to upgrade. & 
        Centralized architecture allows some resource sharing, but still vendor-controlled. & 
        Runs on general-purpose hardware, improving flexibility. & 
        Fully decoupled, allowing software to run on diverse hardware platforms. \\
        \addlinespace
        
        Deployment Cost \& Supply Chain & 
        Expensive hardware, limited supply chain, high upgrade costs. & 
        Shared BBU resources reduce site costs, but core equipment remains expensive. & 
        Reduces capital expenditures by using IT servers, but software remains vendor-controlled. & 
        Competition reduces costs, open ecosystem enhances supply chain resilience. \\
        \addlinespace
        
        Security & 
        End-to-end closed architecture, smaller attack surface, but vendor-controlled security. & 
        Centralized architecture reduces physical attack risks but still depends on a single vendor. & 
        Introduces cloud security risks, increasing virtualization attack surface. & 
        Open interfaces increase attack surface but modular design improves threat response. \\
        \addlinespace
        
        Network Management Model & 
        Managed through a vendor-specific proprietary system. & 
        Unified management of centralized BBU but still vendor-dependent. & 
        Requires integration of IT and telecom management, increasing operational complexity. & 
        Uses open orchestration and RIC for greater flexibility. \\
        \addlinespace
        
        Adaptability to New Technologies & 
        Depends on vendor upgrades, slower adaptation to innovations. & 
        Centralized architecture accelerates new feature deployment. & 
        Supports software-based updates, but vendor reliance remains. & 
        Open ecosystem allows third parties to introduce innovations quickly. \\
        \addlinespace
        
        Ecosystem Openness & 
        Dominated by a few major vendors, limited innovation. & 
        Allows some third-party front-haul solutions, but core components remain closed. & 
        Expands vendor participation, but software ecosystem still restricted. & 
        Fully open standards foster global vendor collaboration and innovation. \\
        
        \bottomrule
    \end{tabularx}
\end{table*}

O-RAN also offers opportunities for enhanced security and privacy protection. Architectural flexibility allows for rapid isolation and mitigation of identified vulnerabilities, thereby reducing the risk of widespread network compromise. In addition, the open and modular nature of O-RAN facilitates the adoption of Continuous Integration/Continuous Deployment (CI/CD) security practices, enabling more dynamic and responsive threat management \cite{bobrovskis2018survey}. In terms of intelligent defence, whilst traditional RAN is also gradually leveraging embedded AI to optimise performance, these capabilities are typically locked into proprietary vendor hardware, making it difficult for operators to intervene. In contrast, O-RAN enables operators or third-party security vendors to deploy customised AI/ML models (such as xApps) via standardised RAN Intelligent Controllers (RICs) and universal computing platforms. These models perform real-time analysis of specific traffic patterns and anomalies. Consequently, O-RAN systems can detect and mitigate complex network threats in a software-defined and modular manner, freeing operators from passive reliance on single-vendor security policy updates \cite{Bonati2021IntelligenceNetworks}. In addition, the modular nature of O-RAN allows for more flexible and dynamic security policies. Operators can update and patch individual components without the need for a complete system overhaul, ensuring that security measures can evolve quickly to address emerging threats.

While O-RAN offers benefits such as enhanced interoperability, reduced vendor lock-in, and the promotion of a more competitive and diverse vendor ecosystem, it also presents a number of security and privacy challenges. The openness of O-RAN increases its exposure to potential cyber threats and vulnerabilities. The integration of artificial intelligence and machine learning technologies, while beneficial for network optimization and resource management, further complicates the security landscape, such as AI-based attacks or flaws in currently unexplained AI \cite{Adadi2018PeekingXAI}. Adversaries can use these systems to disrupt network operations through sophisticated cyberattacks, including denial of service \cite{masdari2016survey}, spoofing \cite{van2018classification} and malicious data injection \cite{illiano2015detecting}. They pose a real risk to critical national infrastructures interconnected through 5G and future networks. Attacks on telecom backbones could lead to catastrophic consequences, such as the disruption of smart city services, industrial operations and essential utilities, illustrating the potential for national security implications. In addition, the distributed nature of O-RAN control functions expands the attack surface, making it imperative that the research and industry communities collaborate to develop and implement robust security measures \cite{Mimran2022EvaluatingNetworks}. This includes standardizing security protocols across all components of the O-RAN architecture and ensuring that these measures mitigate the unique threats posed by the adoption of artificial intelligence and virtualization technologies in the network.

\subsection{Motivation}
Despite these advantages, O-RAN security remains under-researched and requires thorough examination and proactive measures to guard against potential vulnerabilities and privacy breaches. Security is not just a defensive mechanism in O-RAN; it is a crucial safeguard for realizing its key advantages, such as interoperability, vendor diversity, and flexibility. Without a robust security framework, the openness and software-centric features of O-RAN could inadvertently expose the network to countless cyber threats. This could lead to unauthorized access, network operation interruptions, and even complete denial of service, thereby undermining the advantages it aims to provide. Especially with the development of 6G, it is essential to rigorously assess the O-RAN security landscape to address existing and emerging challenges.

O-RAN's dynamic and complex ecosystem, characterized by its reliance on software-defined networking and virtualized network functions, introduces new types of security vulnerabilities that are different from traditional RAN systems. These vulnerabilities not only compromise the integrity and confidentiality of network operations, but also put end-user privacy and data security at risk. The integration of AI and machine learning into network management, while benefiting operational efficiency, further complicates the security matrix by introducing AI-specific threats such as adversarial attacks. Therefore, analyzing security vulnerabilities in O-RANs and identifying robust mitigation strategies are critical steps to ensure the resilience and reliability of future 6G networks. This survey aims to fill the existing research gap by comprehensively analyzing the security challenges of O-RANs, evaluating current and emerging threats, and proposing effective countermeasures to anticipate and eliminate potential risks in the evolving 6G environment.

\subsection{Contribution}

The contribution of this survey to the field of telecom security, especially in the context of O-RAN and the upcoming 6G networks, can be summarized as follows:

\begin{enumerate}

 \item Systematic Classification of O-RAN Security Threats: The survey provides a detailed categorization of security threats specific to the O-RAN architecture, classifying them according to their origin, attack methodology and potential impact. This categorization provides a clearer understanding of the security landscape and helps to develop targeted defense strategies.

 \item Analysis of New Challenges Posed by 6G in O-RAN: An in-depth examination of the new challenges that O-RAN may face with the evolution to 6G, considering the increased complexity and functionality of future networks. This analysis includes potential vulnerabilities introduced by advanced technologies such as non-terrestrial network, Quantum Communications and Terahertz.

 \item Assessment of Existing and Potential Solutions: This survey evaluates current security measures within the O-RAN framework and presents potential solutions to address identified vulnerabilities. The assessment not only covers technical solutions, but also considers the policy, regulatory, and standardization efforts required to effectively protect O-RAN deployments.

 \item We summarize major security reports using a risk-coverage matrix and clarify inherited vs.\ novel risks by contrasting critical 5G interfaces with O-RAN open interfaces.

\end{enumerate}

In addition, Table~\ref{tab:survey_comparison} highlights how this survey differs from previous works that primarily focus on security isolation. Unlike prior studies, this survey systematically integrates 6G-specific challenges, privacy considerations, and mitigation strategies within the O-RAN security landscape. By providing a more holistic perspective, this work serves as a comprehensive reference for researchers and industry practitioners seeking to address the evolving security concerns in O-RAN deployments. By addressing these areas, the survey aims to provide a comprehensive overview of the O-RAN security landscape and identify forward-looking strategies to mitigate the risks associated with the transition to 6G networks.

Table~\ref{tab:survey_comparison} employs a comparative approach using coverage grading to systematically summarise and contrast the content of representative reviews and reports across four dimensions. This methodology enhances interpretability by employing predefined grading criteria to characterise the scope of coverage and the strength of argumentation for studies on specific topics. The specific assessment focuses on four key elements: whether security threat analysis is structured and mappable to O-RAN critical components and interfaces; whether privacy challenges identify sensitive data and data flow boundaries while outlining potential leakage pathways; whether security measures propose implementable controls and discuss deployment constraints and limitations; and whether standardisation discussions conduct substantive analyses of specification requirements versus implementation gaps.

\begin{table*}[t]
\centering
\caption{Comparison of O-RAN security and privacy surveys}
\label{tab:survey_comparison}

\scriptsize
\renewcommand{\arraystretch}{1.25}
\setlength{\tabcolsep}{3pt}

\newcolumntype{Y}{>{\raggedright\arraybackslash}X}

\begin{tabularx}{\textwidth}{@{} l Y Y Y Y Y @{}}
\toprule
\textbf{Article} &
\textbf{Security threat analysis} &
\textbf{Privacy challenges} &
\textbf{Security measures evaluation} &
\textbf{Standardization discussion} &
\textbf{Summary} \\
\midrule

\cite{BSI5GRAN2023} &
\textbf{Strong}---Methodical O-RAN risk analysis with clear scope; limited empirical grounding. &
\textbf{Moderate}---Identifies privacy-sensitive data exposure paths; limited governance/controls depth. &
\textbf{Moderate}---Recommends safeguards, but implementation and evaluation details are limited. &
\textbf{Comprehensive}---Extensive discussion of O-RAN/3GPP assumptions, specification gaps, and deployment implications. &
Security- and standardization-centric risk study; privacy and empirical evidence are secondary. \\
\addlinespace

\cite{OpenRAN2022} &
\textbf{Strong}---Broad cybersecurity risk assessment for Open RAN, including supply-chain and virtualization concerns. &
\textbf{Limited}---Privacy is largely implicit (e.g., CIA(+P) framing) with minimal dedicated analysis. &
\textbf{Moderate}---Provides mitigation directions, but mostly at a high level with limited engineering specifics. &
\textbf{Strong}---Substantive standards/policy discussion (ecosystem roles, transparency, regulatory implications). &
Policy/report-style analysis emphasizing supply chain and governance; limited technical privacy depth. \\
\addlinespace

\cite{CISAOpenRAN2022} &
\textbf{Strong}---Identifies attack surfaces and security considerations across major Open RAN components and interfaces. &
\textbf{Limited}---Privacy is briefly mentioned; limited privacy threat taxonomy or compliance discussion. &
\textbf{Strong}---Actionable security control guidance (interfaces, platform, monitoring), with limited validation. &
\textbf{Moderate}---References ecosystem guidance/specs but limited comparative standardization analysis. &
Practical security considerations and control recommendations; evaluation evidence is limited. \\
\addlinespace

\cite{IFRIOpenRAN2022} &
\textbf{Limited}---Security is framed mainly as strategic/industrial risk rather than a technical threat model. &
\textbf{Not covered}---Privacy is not treated as a primary analytical dimension. &
\textbf{Limited}---Mitigation focuses on policy/industry measures; few concrete technical controls. &
\textbf{Strong}---Deep discussion of standardization politics, stakeholder incentives, and international dynamics. &
Geopolitics-centered perspective; strong on policy/standardization context, light on technical security/privacy. \\
\addlinespace

\cite{NTTDOCOMO5GOpenRAN2023} &
\textbf{Moderate}---Summarizes Open RAN security concerns without a formal threat model. &
\textbf{Not covered}---Little to no explicit privacy analysis. &
\textbf{Moderate}---Reviews practical measures (e.g., lifecycle/operations) but limited technical depth and evaluation. &
\textbf{Moderate}---Mentions ecosystem/interoperability efforts; limited standards/security-option analysis. &
Ecosystem/industry overview; useful deployment context but not a deep security/privacy assessment. \\
\addlinespace

\cite{NTIAOpenRAN2023} &
\textbf{Comprehensive}---Structured threat modeling (e.g., STRIDE) and broad attack-surface coverage. &
\textbf{Strong}---Explicit privacy/confidentiality considerations in cloud-native, multi-vendor Open RAN settings. &
\textbf{Comprehensive}---Maps mitigations to requirements and discusses implementability/testing considerations. &
\textbf{Comprehensive}---Specification-driven analysis with concrete security requirements and deployment implications. &
Deep security report linking threats, mitigations, and standards; strong structure and breadth. \\
\addlinespace

\cite{Liyanage2023OpenOpportunities} &
\textbf{Comprehensive}---Broad taxonomy spanning interface/component, cloud/platform, operations, and AI/ML threats. &
\textbf{Strong}---Covers AI-related privacy leakage risks and data exposure surfaces; mitigation depth varies. &
\textbf{Strong}---Summarizes mitigation approaches and governance practices; limited empirical validation. &
\textbf{Strong}---Covers standards and security frameworks with clear gaps and open issues. &
Wide-coverage academic survey; strong synthesis, with relatively limited evaluation evidence. \\
\addlinespace

\cite{chen2023briefsurveyopenradio} &
\textbf{Moderate}---Brief O-RAN security overview emphasizing open architecture and ML-centric threats. &
\textbf{Limited}---Notes privacy leakage (e.g., inference risks) but limited privacy governance discussion. &
\textbf{Moderate}---Summarizes mitigations at a high level; limited systematic evaluation. &
\textbf{Moderate}---Mentions standardization context but limited depth. &
Concise ML-oriented overview; good entry point but not a comprehensive baseline. \\
\addlinespace

\textbf{This survey} &
\textbf{Comprehensive}---Systematic 6G-oriented threat taxonomy for O-RAN with clear component/interface mapping (incl.\ STRIDE). &
\textbf{Comprehensive}---Covers 6G privacy risks (AI/telemetry, cross-vendor data sharing, regulatory compliance) and implications. &
\textbf{Comprehensive}---Synthesizes mitigation strategies with deployability and overhead considerations, supported by empirical evidence reported in prior studies. &
\textbf{Comprehensive}---Surveys O-RAN security specifications and regional policy frameworks; highlights optionality and alignment gaps for 6G. &
Holistic 6G-focused synthesis of O-RAN security, privacy, mitigation, and standardization. \\

\bottomrule
\end{tabularx}

\begin{tablenotes}
    \item \textbf{Comprehensive} - Structured, clear O-RAN scenario mapping, with implementation or constraints or validation discussion. \\
    \item \textbf{Strong} - Structured and fairly specific, but lacks implementation details or validation support or privacy governance depth. \\
    \item \textbf{Moderate} - Covers key points partially, mostly high-level, with limited technical substantiation. \\
    \item \textbf{Limited} - Mentions only briefly, narrow coverage, lacking systematic analysis. \\
    \item \textbf{Not covered} - Essentially absent, or only generic statements.
\end{tablenotes}
\end{table*}

\subsection{Outline}

\begin{figure*}[htbp]
\centering
\includegraphics[width=7in]{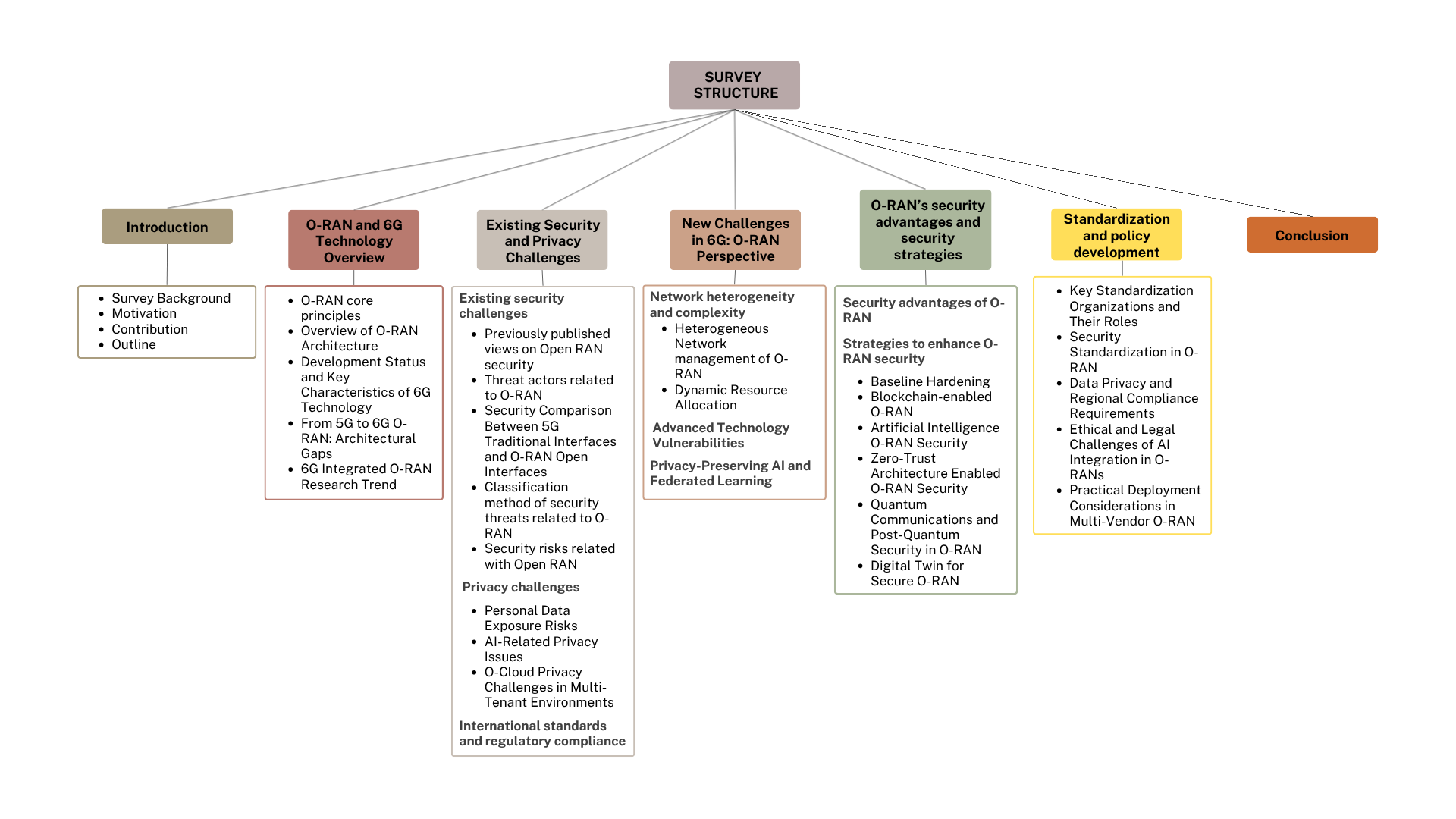}
\caption{Organization structure of this paper}
\label{Structure}
\end{figure*}

This paper is organized as shown in Fig. \ref{Structure}: Chapter II details the O-RAN architecture and its characteristics, while exploring the concept of 6G networks and the expected development trends, and analyzing how O-RAN can be integrated with the upcoming 6G network technologies. Next, Chapter III focuses on the current security and privacy issues facing O-RANs, by analyzing the threat surface of O-RANs in depth and classifying these security threats using the STRIDE model. Chapter IV predicts new security and privacy challenges that O-RAN may encounter in the 6G environment, especially in the emerging technology areas such as Space-Air-Ground integrated networks and quantum communications. Chapter V describes the security benefits of O-RAN and its best practices, and explores which technologies can enhance the security applications of O-RAN in 6G networks. Section VI summarizes the current standards and regulations for O-RAN in terms of security and privacy, providing the reader with a comprehensive view of O-RAN security.

\section{O-RAN and 6G technology overview}
\subsection{O-RAN core principles}

O-RAN is a major innovation to traditional RANs, introducing standardized interfaces and an open, flexible and intelligent network architecture. Traditional RANs are proprietary systems where all components from hardware to software are tightly integrated and supplied by a single vendor. This monopolistic approach limits interoperability and flexibility as mobile network operators (MNOs) are constrained to vendor-specific technologies and upgrades, leading to higher costs and less innovation due to vendor lock-in. Unlike previous closed systems from a single vendor, O-RAN emphasizes interoperability and standardization of RAN components, including hardware and software from different vendors, to work together seamlessly. And through Software Defined Networking (SDN) \cite{Niknam2022IntelligentNetworks} and Network Functions Virtualization (NFV) \cite{Garcia-Saavedra2021O-RAN:Ecosystem} technologies, the O-RAN architecture allows for a modular approach to integrating base station software into off-the-shelf hardware, enabling network operators to rapidly deploy and scale services. Since the inception of the O-RAN Alliance in 2018, the concept of ``open and interoperable'' has been widely recognized and supported by the industry \cite{xRANForum2018XRANAlliance}. In comparison to traditional RAN structures, O-RAN has several key principles and features \cite{Polese2023UnderstandingChallenges}:

\subsubsection{Decomposition}
\begin{figure}[!t]
\centering
\includegraphics[width=3.5in]{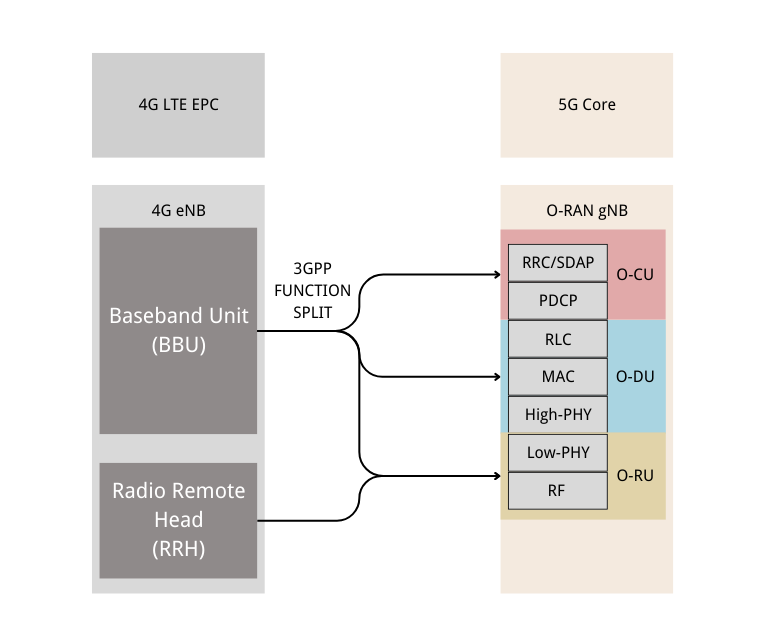}
\caption{Function migration from 4G to 5G}
\label{Function migration from 4G to 5G}
\end{figure}

Functional splits in O-RAN refer to how the different processing tasks of a base station are divided among its components (i.e., RUs, DUs, and CUs). This division determines the extent to which baseband processing is handled locally (close to the user) and centrally (in a data centre or cloud environment). The location of the division within the protocol stack affects factors such as latency, bandwidth requirements, and computational load distribution. The O-RAN Alliance has a flexible architecture that supports multiple functional split options, allowing network operators to customise deployments based on network conditions and service requirements. One of the most commonly used splits is the 7.2x split, which divides physical layer processing between RUs and DUs. This approach helps optimise the fronthaul bandwidth while maintaining network performance \cite{Larsen2019A}. As shown in Fig. \ref{Function migration from 4G to 5G}, this particular split divides the traditional Base Band Unit (BBU) into three distinct components: CU, DU and RU. This configuration is just one example of how O-RAN can modularize network functions to enhance management efficiency and support diverse deployment scenarios. The CU is further divided into two logical components that deal with the control plane (CP) and user plane (UP). The RU handles the digital front end (DFE), some of the physical layer functions, and the digital beamforming functions, while the DU is responsible for the rest of the physical layer, the media access control (MAC) layer, and the radio link control (RLC) layer. These functions include scrambling, modulation, layer mapping and partial precoding, as well as mapping to the Physical Resource Block (PRB). The CU unit is responsible for implementing the higher level functions of the 3GPP protocol stack, including the Radio Resource Control (RRC) layer, the Service Data Adaptation Protocol (SDAP) layer and the Packet Data Convergence Protocol (PDCP) layer. These functions include managing the quality of service of traffic, encryption of data, etc.

In traditional RAN architecture, the BBU takes on all these functions, and processing from the physical layer to the network layer is done in a centralized unit. O-RAN enables these functions to be distributed in different units by introducing split 7.2x. In addition, split 7.2x specifically focuses on keeping some of the physical layer functions in the RUs and distributing the rest to the DUs and CUs. this split provides advantages in reducing the complexity and power consumption of the RUs, simplifying network upgrades and maintenance, and supporting multiple radio access technologies. Also, this segmentation enables the Ethernet-based eCPRI protocol, which supports low-latency and cost-effective forward connections.

\subsubsection{RAN Intelligent Controller and Closed-Loop Control}
The second feature of O-RAN involves the RAN Intelligent Controller (RIC) and closed-loop control. RIC, as a programmable component, achieves closed-loop control by running optimization programs and coordinating the RAN. O-RAN architecture includes two different types of RIC: non-real-time RIC and near-real-time RIC, both of which utilize advanced AI and ML algorithms to achieve dynamic and intelligent optimization of RAN functions. The non-real-time RIC (Non-RT RIC) interacts with the network orchestrator, operating over a time scale of more than one second, helping to execute tasks such as policy management and long-term resource optimization. In contrast, near-real-time RIC (Near-RT RIC) coordinates delay-sensitive control loops within RAN nodes, operating on a fine time scale from 10 ms to 1 second, thereby supporting critical real-time decision-making processes.

In 4G networks, dynamic resource allocation and network optimization usually rely on the capabilities of the network equipment itself and the operator's network management system. For example, for large events such as sports events, operators may plan ahead and manually adjust cell configurations (e.g., increasing the bandwidth allocation of cells, adjusting power output, etc.) to cope with expected high traffic demand. However, these adjustments often require manual intervention and have slow response times. O-RAN enables the network to manage and optimize resources with an unprecedented level of flexibility and intelligence by integrating RICs. RIC supports finer-grained network analysis and real-time decision making, and is able to dynamically adjust resource allocation based on immediate network conditions and traffic data. For example, in the same sports event scenario, with the intervention of near real-time RIC, the network can monitor the specific traffic situation in the user-intensive area in real time and automatically adjust the resource allocation, such as dynamically adjusting the power of the neighboring cells or adjusting the cells to which the users are connected in real time, in order to balance the load of the network and improve the user experience.

\subsubsection{Virtualization}
The third principle of the O-RAN architecture is virtualization, which allows the introduction of additional components or software (xApps) to manage and optimize network infrastructure and operations from the edge system to the virtualized platform. And all O-RAN components except O-RU can be deployed on a hybrid cloud computing platform called O-Cloud, combining physical nodes, software components, and management and orchestration capabilities. O-Cloud decouples hardware and software and supports hardware sharing between different tenants. For example, virtualization enables network slicing, allowing operators to create multiple virtual networks on the same physical network infrastructure, each with its own specific quality of service and characteristics. This supports customized services for different user groups and applications, such as low-latency, high-reliability services. In addition, virtualization capabilities allow operators to easily and dynamically scale up or down the computing resources needed to support subscriber demand, rather than planning computing resources based on manual effort, which limits the power consumption of the actual network functions required \cite{garcia2021nuberu} \cite{sabella2014energy}.

\subsubsection{Open Interfaces}
O-RAN breaks down vendor lock-in in the RAN through technical specifications and by defining open interfaces for communication between a variety of different components, thereby increasing market competitiveness, innovation, and faster update and upgrade cycles. These standardized interfaces allow components from different vendors to interact, e.g., a Non-RT RIC from one vendor can interact with a base station from another vendor, thus facilitating interoperability between CUs, DUs, and RUs from different manufacturers. The following is a brief overview of some of the key interfaces and their protocols in the O-RAN:
\begin{itemize}
    \item \textbf{F1 interface:} Connects CUs and DUs. The protocols supported by the F1 interface include the F1 application layer protocols for transferring control signaling and user data. The F1 interface allows communication between CUs and DUs from different vendors.

    \item \textbf{E2 interface:} Connects the Near-RT RIC to network functions (e.g., DUs and CUs).The protocols supported by the E2 interface are designed to enable the RIC to manage and optimize RAN behavior through control and optimization policies. The E2 interface supports a specific set of service models that define the types of information exchanged between the RIC and network functions.

    \item \textbf{A1 interface:} Connects the Non-RT RIC to the Near-RT RIC. The protocol of the A1 interface supports the exchange of policies and models to guide the operational and optimization decisions of the near real-time RIC.

    \item \textbf{O1 Interface:} For communication between Management and Orchestration (MANO) and network functions.The protocols supported by the O1 interface allow network operations and management systems to perform resource management, performance monitoring, and fault management in a standardized manner.

    \item \textbf{O2 Interface:} For communication between cloud resource managers (e.g., Virtualized Infrastructure Manager, VIM) and Service Management and Orchestration (SMO). The protocols and models supported by the O2 interface are designed to optimize resource allocation and network service deployment.
\end{itemize}

\subsection{Overview of O-RAN Architecture}

\begin{figure}
\centering
\includegraphics[width=3.5in]{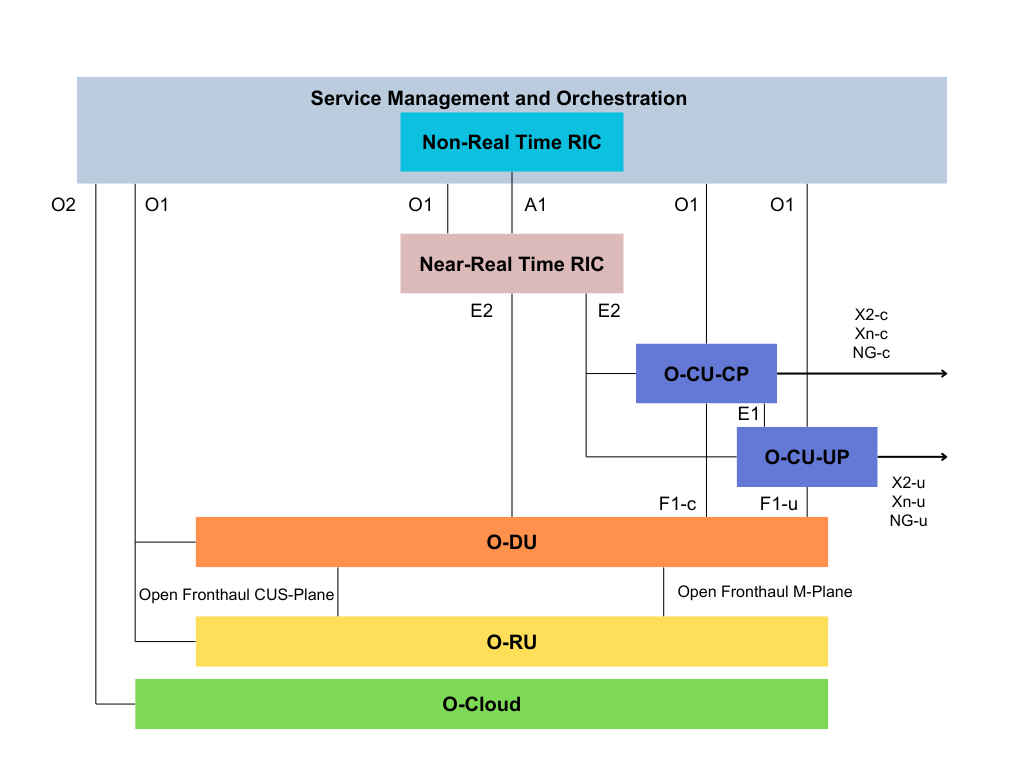}
\caption{O-RAN logical architecture}
\label{Architecture}
\end{figure}

The main elements of O-RAN's architecture include O-RAN Central Unit (O-CU), O-RAN Distributed Unit (O-DU), O-RAN Radio Unit (O-RU), RIC, SMO, and O-Cloud. As shown in Fig. \ref{Architecture}.
\begin{itemize}
    \item \textbf{O-CU:} A software-based central unit that handles higher-level RAN protocols. Logically divided into control-plane and user-plane components, O-CU is responsible for functions such as mobility management, session management, and handling of RAN sliced traffic. It communicates with O-DUs through an open F1 interface and typically operates in a centralised or cloud environment to aggregate traffic from multiple DUs.

    \item \textbf{O-DU:} A software-driven distributed unit for baseband processing functions closer to the physical layer (e.g., scheduling, error correction coding, ARQ). O-DUs are connected to one or more O-RUs via an Open Fronthaul Interface (usually 7.2x split via the Open Fronthaul Standard) and to O-CUs via F1. By virtualising the DUs on a cloud infrastructure at the edge of the network, the O-RAN allows for dynamic scaling of processing resources in response to changes in demand.

    \item \textbf{O-RU:} A radio unit that includes an RF front-end and RF processing components (antenna, amplifier, converter) as well as some physical layer processing. O-RU is usually located at the base station (antenna location) and is responsible for transmitting and receiving radio signals over the air interface. It interacts with the rest of the RAN through a standardised management plane (M-plane) and a forward transmission interface, allowing centralised control and configuration from the SMO, while processing real-time RF functions locally.

    \item \textbf{RIC:} RIC is the cornerstone of the O-RAN architecture, introducing a platform for hosting RAN control applications. As part of the SMO framework, Non-RT RIC runs at control loop times of seconds or longer and hosts rApps (resource management applications) that perform policy-based control, global optimisation and provide guidance to the near-real-time layer. Near-RT RIC runs on millisecond control loops and hosts xApps that perform localised, low-latency optimisation tasks (e.g. real-time load balancing, disturbance mitigation or switching optimisation). The rApps and xApps work together to enable AI-driven and closed-loop control at different timescales of the RAN.

    \item \textbf{SMO:} Responsible for the overall management of O-RAN components and resources. It provides orchestration of the RAN VNF/CNF, handles FCAPS management (fault, configuration, accounting, performance, security), and hosts non-RT RIC functions. SMO communicates with RAN nodes through O-RAN's standard interfaces (A1, O1, O2, etc.).

    \item \textbf{O-Cloud:} A cloud infrastructure (data centre or edge cloud) hosting O-RAN software functions. O-Cloud provides a virtualised environment for O-CU, O-DU and RIC components. It is essentially a pool of compute, storage and network resources managed by a cloud platform such as an NFV infrastructure manager or Kubernetes for CNF. O2 interfaces connect SMOs to O-Cloud's management system to allocate resources on demand and deploy network functions as needed. By abstracting hardware, O-Cloud enables the RAN to benefit from the scalability and reliability of the cloud and to quickly deploy updated or new network slices as software instances.
\end{itemize}
Additionally, the O-RAN Alliance defines various different interfaces, including A1, O1, E1, F1, open fronthaul M-plane, and O2, to facilitate flexible connections and collaboration among different components.

\subsection{Development Status and Key Characteristics of 6G Technology}

Although global 5G networks are still in the early stages of large-scale deployment, academia and industry have already begun actively exploring and defining the technical standards and development pathways for 6G. According to the International Telecommunication Union (ITU)'s preliminary planning, 6G networks are expected to achieve initial commercial deployment around 2030, and related standard-setting efforts (such as the ITU-R's IMT-2030 framework) have already been launched \cite{InternationalTelecommunicationUnionRadiocommunicationSector2023IAFIFinal}. Compared to 5G, 6G will exhibit even more extreme performance metrics. A detailed quantitative comparison of the Key Performance Indicators (KPIs) between 5G (IMT-2020) \cite{ITU-R_M.2083-0} and 6G (IMT-2030) \cite{ITU-R_M.2160-0} is presented in Table \ref{tab:5g_vs_6g}.

\begin{table}[htbp]
\centering
\caption{Comparison of Key Performance Indicators (KPIs) between 5G and 6G}
\label{tab:5g_vs_6g}
\resizebox{\columnwidth}{!}{
\begin{tabular}{@{}lll@{}}
\toprule
\textbf{Key Performance Indicator} & \textbf{5G (IMT-2020)} & \textbf{6G (IMT-2030 Vision)} \\ \midrule
\textbf{Peak Data Rate} & 20 Gbps & 50--200 Gbps (up to 1 Tbps) \\
\textbf{User Experienced Data Rate} & 100 Mbps & 300--500 Mbps \\
\textbf{Latency (Air Interface)} & 1 ms & 0.1--1 ms \\
\textbf{Connection Density} & $10^6$ devices/km$^2$ & $10^7$--$10^8$ devices/km$^2$ \\
\textbf{Mobility} & 500 km/h & 500--1000 km/h \\
\textbf{Spectrum Efficiency} & 3$\times$ vs. IMT-Advanced & 1.5$\times$--3$\times$ vs. IMT-2020 \\
\textbf{Positioning Accuracy} & N/A (Enhanced later) & 1--10 cm (Indoor) \\
\bottomrule
\end{tabular}%
}
\end{table}

From a technical perspective, the three key features of 6G are: comprehensive network intelligence, seamless coverage across all areas, and the deep integration of communication and multi-dimensional functions:

\begin{itemize}
    \item \textbf{Comprehensive Intelligence: }5G era has continued and enhanced the self-organising network (SON) \cite{moysen20184g} mechanism introduced in 4G/LTE to achieve automatic network configuration, self-optimisation, and self-healing. However, 6G network intelligence will evolve into a higher level of autonomous networks capable of long-term self-sustainability and self-adaptation. By deploying large-scale AI algorithms, the network not only possesses real-time dynamic resource allocation and predictive load balancing capabilities but can also proactively sense its own operational status, quickly locate faults, actively repair them, and perform closed-loop optimisation. Additionally, 6G networks will leverage machine learning and deep reinforcement learning technologies to achieve more precise energy management and long-term self-evolution, adapting to evolving business requirements and environmental conditions \cite{zhang20196g}.

    \item \textbf{Seamless Global Coverage: }Traditional terrestrial cellular networks are limited by geographical conditions and cannot cover all areas. 6G aims to break through the limitations of terrestrial networks by leveraging a space-ground integrated network (SAGIN) \cite{8368236}. Through a collaborative networking approach involving satellites, high-altitude platforms (such as high-altitude balloons or drones), and ground base stations, 6G will provide global users with seamless, stable 3D coverage. This network will for the first time enable continuous, high-speed, and stable connections for users in mobile scenarios such as aircraft, high-speed trains, and maritime transportation, and effectively address communication needs in special environments such as remote areas and disaster sites.
    
    \item \textbf{Deep Integration of Communication, Computing, Control, Localization, and Sensing (3CLS):}Unlike previous network systems that focused on data communication, 6G network nodes will simultaneously perform multiple functions, including data transmission, real-time computing, environmental sensing, precise positioning, and remote control. It supports the real-time, high-precision interaction requirements of emerging application scenarios such as smart manufacturing, autonomous driving, and augmented reality \cite{saad2019vision}.
\end{itemize}

\subsection{From 5G to 6G O-RAN: Architectural Gaps}
    Although current O-RAN architecture solves traditional RAN vendor lock-in and flexibility problems through decoupling and the adopting RIC, its performance metrics are presently primarily based on 5G standards such as 3GPP Release 15/16 \cite{3gpp_rel15} \cite{3gpp_rel16}. Facing the 6G network vision, the existing O-RAN architecture still exhibits evolutionary gaps in real-time capabilities, support for heterogeneous networks, and native intelligence. To meet 6G requirements, O-RAN needs to make profound evolutions across the following key architectural dimensions. Table~\ref{tab:oran_evolution_comparison} summarizes the key architectural shifts as O-RAN evolves from a 5G-oriented design toward 6G requirements along the following dimensions.

    \begin{itemize}
        \item \textbf{Intelligent Evolution:} Current O-RAN primarily relies on Non-RT RIC and Near-RT RIC to execute optimisation tasks, with control loop time scales typically ranging from 10 ms to 1 second or longer. While this design is acceptable for traffic control in the 5G era, typical 6G application scenarios-such as holographic communication and Haptic Internet-demand ultra-low latency response capabilities from the network. To fulfill this gap, future O-RAN may evolve towards an AI-Native architecture \cite{li2024architecture}. This would involve introducing real-time intelligent control modules co-located with O-DU \cite{ko2023edgeric}, shifting part of the RIC's inference capabilities to the edge. This enables real-time intelligent decision-making, rather than solely relying on asynchronous optimisation from upper-layer controllers.
    
        \item \textbf{Coverage Expansion:} Current O-RAN Open Fronthaul is primarily designed for terrestrial cellular network scenarios, assuming relatively stable fibre connections and low-latency transmission environments. However, 6G aims to integrate satellites, high-altitude platform systems (HAPS), and UAV base stations to achieve seamless global coverage. The high latency, dynamic topology changes, and Doppler shift characteristics of satellite links make existing synchronisation mechanisms and standard interface protocols (such as eCPRI) difficult to directly reuse \cite{10188308} \cite{seeram2025handover}. Consequently, O-RAN architecture must expand its functional split options to fit the constrained computational capabilities and unstable backhaul links of non-terrestrial nodes. This includes exploring flexible deployment of lightweight O-DU or O-CU components on satellite payloads.
    
        \item \textbf{Performance Enhancement:} O-RAN emphasises running vRAN on commercial off-the-shelf (COTS) hardware, offering deployment flexibility in the 5G era. However, the anticipated peak data rates for 6G-ranging from 50 Gbps to 200 Gbps, will impose significant processing pressure and energy consumption burdens on general-purpose CPUs. Relying merely on software processing will encounter severe throughput bottlenecks \cite{park2025accelerating}. Therefore, the evolution of 6G O-RAN needs to incorporate dedicated hardware accelerators (such as FPGAs, ASICs, or GPU-based acceleration) to offload computational-intensive physical layer processing \cite{schiavo2024cloudric}. How to achieve hardware acceleration while maintaining interface openness will become one of the key challenges in integrating the 6G O-RAN architecture.
    \end{itemize}
    
    \begin{table}[t]
        \centering
        \caption{Architectural contrast between 5G-oriented O-RAN deployments and 6G O-RAN targets}
        \label{tab:oran_evolution_comparison}
        \scriptsize
        \begin{tabularx}{\linewidth}{@{}
            >{\bfseries}l
            >{\raggedright\arraybackslash}X
            >{\raggedright\arraybackslash}X
        @{}}
        \toprule
        \textbf{Dimension} & \textbf{5G O-RAN Baseline} & \textbf{6G O-RAN Target} \\
        \midrule
        
        \textbf{AI Architecture} &
        Add-on AI functions (xApps/rApps) &
        AI-native \\[0.35em]
        
        \textbf{Inference Placement} &
        Centralized at RIC (Cloud/Edge Cloud) & 
        Hierarchically distributed across RIC and O-DU \\[0.35em]
        
        \textbf{Network Topology} &
        Primarily 2D terrestrial macro/micro-cell deployments &
        3D space-air-ground integrated (SAGIN) topology \\[0.35em]
        
        \textbf{Compute Platform} &
        General COTS (Software-centric) & 
        Heterogeneous (CPU + FPGA/GPU/ASIC) \\[0.35em]
        
        \textbf{Peak Throughput} &
        $\approx$ 20 Gbps (eMBB Standard) & 
        50 Gbps -- 200 Gbps \\
        
        \bottomrule
        \end{tabularx}
    \end{table}

    Overall, this architectural evolution-featuring more distributed edge intelligence nodes and a more complex SAGIN topology-brings not only enhanced performance metrics but also changes the system's trust boundaries and attack surface. Existing security models (to be discussed in Chapter 3) are primarily designed for 5G terrestrial networks and relatively static cloud environments, proving inadequate against novel 6G threats such as adversarial sample attacks targeting real-time AI or physical capture risks affecting satellite nodes. This architectural evolution introduces entirely new security challenges, which Chapter 4 will discuss in depth.

\subsection{6G Integrated O-RAN Research Trend}
Several research and standardisation efforts are already underway to integrate O-RAN with 6G. We highlight some noteworthy trends and research that illustrate how to shape O-RAN in the 6G era:

\begin{itemize}
    \item \textbf{Industry Initiative:} The O-RAN Alliance's Next Generation Research Group (nGRG) \cite{oran_introduction_2023} has been established to explore the principles of open and intelligent RAN in the context of 6G. The group is developing a 6G research agenda and key priorities for O-RAN, with a focus on ensuring a smooth and sustainable transition from 4G/5G to 6G networks. It also aims to unify the technology path for 6G to avoid incompatibility with other standards development organisations (SDOs).
    \item \textbf{O-RAN Architecture Analysis:} In the academic field, there are active researches on how O-RAN can cope with the 6G challenges. The authors of \cite{Abdalla2022TowardDo} provide a comprehensive analysis of the strengths and limitations of the O-RAN architecture, with a particular focus on its readiness for future networks. Notably, their study included a survey of 95 researchers from around the world, and the vast majority agreed that O-RAN will be the cornerstone of next-generation (6G) network infrastructure. The article concludes with a discussion of O-RAN's strengths (e.g., flexibility and support for AI) as well as current weaknesses and unresolved issues.
    \item \textbf{RAN Evolution and 6G Functionality:} Other research efforts have investigated how the RAN will evolve in the 6G era and how O-RAN will fit into this evolution. Studies such as \cite{Diego2020EvolutionNetwork} and \cite{Singh2020TheOpportunities} discuss the gradual shift in traditional RAN functionality and implementation to the O-RAN model, showing that many of the concepts adopted in the post-5G era (e.g., NFV, disaggregated RAN nodes, cloud-native network management) are stepping stones to the eventual realisation of a fully open and intelligent 6G RAN. In particular, a set of studies \cite{Fattore2020AutoMEC:Resources}\cite{Brik2020Service-OrientedArchitecture}\cite{Chih-Lin2020AEra}\cite{Brik2021TowardApproach} delve into how O-RANs can support functions beyond 5G such as network slicing and Mobile Edge Computing (MEC) in a multi-vendor environment. These works demonstrate the potential of O-RAN as a new 6G service infrastructure. At the same time, they acknowledge that challenges remain, such as the need for rigorous interoperability testing, real-time performance optimisation in disaggregated environments, and ensuring reliability and QoS across slices. The consensus that emerges from these research trends is that combining O-RAN with 6G is not only feasible, but is already underway.
\end{itemize}

\section{Existing Security and Privacy Challenges in O-RAN}
\subsection{Existing Security Challenges}
\subsubsection{Previously Published Views on O-RAN Security}

With the rise of O-RAN technology in the global communications industry, its security and privacy implications and challenges are coming into focus. While O-RAN aims to increase network flexibility and efficiency through open and virtualized network elements, this new architecture also poses unique security threats. In order to comprehensively understand and address these challenges, several international organizations and institutions have released a series of analytical reports on O-RAN security. The reports in Table \ref{tab:openran_matrix} mainly analyse the potential security risks faced by O-RAN and propose some security countermeasures and recommendations.

\begin{table*}[t]
\footnotesize
\centering
\begin{threeparttable}
\caption{Risk-coverage matrix for major O-RAN security reports}
\label{tab:openran_matrix}
\begin{tabularx}{\textwidth}{l*{7}{>{\centering\arraybackslash}p{0.55cm}}X}
\toprule
\textbf{Report} & \textbf{SPEC} & \textbf{OSS} & \textbf{CLOUD} & \textbf{SC} & \textbf{AS} & \textbf{LC} & \textbf{TEST} & \textbf{Summary}\\
\midrule
BSI \cite{BSI5GRAN2023}            & 2 & 0 & 2 & 1 & 2 & 2 & 0 & It qualitatively assesses Open RAN risks under best- vs worst-case security implementations. It finds O-RAN specifications are not sufficiently ``secure by design'' - many controls are optional or ambiguous - leading to multiple medium/high risks across interfaces (Open Fronthaul, O-Cloud, etc.) and calls for mandatory security measures in the standards.\\
NIS Group \cite{OpenRAN2022}      & 1 & 2 & 2 & 2 & 2 & 1 & 1 & It examines how O-RAN impacts 5G security risks and opportunities. It notes that O-RAN's new paradigm (new interfaces, multi-vendor components) expands the attack surface and introduces uncertainties (e.g. reliance on cloud providers), while also offering potential benefits (vendor diversity, transparency) if supporting measures like mature standards and EU 5G Toolbox mitigations are applied.\\
CISA \cite{CISAOpenRAN2022}     & 0 & 2 & 2 & 1 & 2 & 2 & 0 & It highlights four O-RAN security considerations: multi-vendor management complexity, Open Fronthaul interface protections (confidentiality, integrity, availability), rApp/xApp supply chain risks, and AI/ML vulnerabilities. It stresses that O-RAN faces familiar software, open-source, and supply chain threats (as in broader ICT) and urges adopting best practices (encrypting all interfaces, mutual authentication for inter-component connections, zero-trust architectures, etc.) to mitigate these issues. \\
IFRI \cite{IFRIOpenRAN2022}           & 1 & 2 & 0 & 2 & 2 & 1 & 0 & It praises advantages like virtualization, automation, and disaggregation (improving flexibility and reducing vendor lock-in) but warns that these same features can increase security risks - a more complex, multi-supplier RAN raises misconfiguration and performance issues, expands the attack surface (new interfaces, increased use of open-source code), and lacks transparency in its standardization process.\\
NTT Docomo \cite{NTTDOCOMO5GOpenRAN2023}     & 0 & 2 & 2 & 0 & 2 & 2 & 0 & It devotes a brief section to security, identifying several challenges. It flags risks such as vulnerabilities in open-source/COTS software, an increased attack surface from disaggregated architecture, functional security gaps in newly introduced RAN elements, greater physical attack exposure (distributed sites), and cloud/virtualization threats, as well as process challenges like higher complexity in multi-vendor operations and the need for comprehensive security lifecycle management.\\
NTIA \cite{NTIAOpenRAN2023}    & 2 & 2 & 2 & 2 & 2 & 2 & 2 & It provides an in-depth comparative risk analysis of O-RAN vs. traditional RAN. While noting potential security benefits of O-RAN, it finds that many risks often attributed to O-RAN (e.g. supply chain threats, software vulnerabilities) also exist in conventional RAN, and it emphasizes addressing O-RAN's unique challenges - such as specification optionality and new inter-vendor interfaces - via stricter standards and holistic, lifecycle-based security measures.\\
O-RAN Alliance \cite{oranSpec2024} & 2 & 2 & 2 & 2 & 2 & 2 & 2 & It delivers a comprehensive threat model and risk assessment for O-RAN. It catalogues a wide range of threats across the O-RAN system (from O-Cloud and virtualization vulnerabilities to open-source code risks, physical attacks, and protocol/API exploits) and defines corresponding security controls/principles (e.g. mutual authentication, access control, secure boot, robust isolation, secure update) to mitigate these risks and strengthen O-RAN security-by-design.\\
\bottomrule
\end{tabularx}

\begin{tablenotes}[flushleft]\footnotesize
\item \textbf{Scores}: 0 = not covered; 1 = mentioned; 2 = analysed in depth.
\item \textbf{Dimensions}: SPEC = Specification gaps/optionality; OSS = Open-source software risk; CLOUD = Virtualization \& cloud security; SC = Supply-chain security; AS = Attack surface \& complexity; LC = Lifecycle / governance / IAM; TEST = Security testing \& certification.
\end{tablenotes}
\end{threeparttable}
\end{table*}

Although reports from BSI, NTIA, CISA, and multiple O-RAN security reviews have established systematic risk landscapes and hardening recommendations, these are mostly based on current 5G deployments. Several studies indicate that traditional approaches - such as strong encryption, frequent mutual authentication, and multi-hop tunnel stacking - may significantly amplify control loop latency and jitter, thereby directly affecting the reliability of latency-sensitive services like URLLC and collaborative sensing \cite{abdel2022security}. Existing O-RAN security reports generally recommend enabling encryption and zero-trust authentication on interfaces such as the open fronthaul, A1/E2, and O1/O2. However, they seldom provide joint performance analyses for critical control plane scenarios like RIC closed-loop control and slice elastic orchestration. Furthermore, they lack a systematic approach incorporating hardware offloading, lightweight cryptography, and protocol tailoring. Directly migrating these recommendations to 6G-O-RAN risks achieving configuration compliance while exhausting the overall latency budget through security mechanisms themselves \cite{saeed2025comprehensive}.

Meanwhile, existing reports lack systematic modelling of novel attack vectors arising from AI control and SAGIN integration within 6G scenarios. Existing frameworks typically categorise these new functionalities simplistically as `high-risk applications' or `high-value assets', without explicitly incorporating these novel attack vectors into threat modelling. Furthermore, a significant number of security controls within current O-RAN specifications remain `recommended/optional'. This leads to substantial implementation variations among vendors regarding interface hardening, logging and auditing, and key lifecycle management, resulting in a significant security gap between specification and deployment.

Overall, existing reports provide an important security baseline for O-RAN, yet they exhibit significant limitations in addressing 6G performance constraints, new attack surfaces such as AI/SAGIN, and cross-vendor enforcement mechanisms.

\subsubsection{Threat Actors Related to O-RAN}

When assessing security risks to the O-RAN, it is critical to identify and classify threat actors based on their interaction with the system. This categorisation helps to understand the various potential risks and develop appropriate security measures. The types of threat actors identified include hardware supply chain, independent software vendors (ISVs), user equipment (UEs), insider threats, and external threats \cite{Mimran2022SecurityNetworks}. Each category represents a different pattern of interaction with the O-RAN system and presents unique challenges ranging from internal vulnerabilities to external attacks that have a significant impact on the security of the network.

\begin{itemize}
    \item \textbf{Hardware Supply Chain}: There are two main components, the entities that produce O-RAN hardware components and the entities involved in the distribution of the components, i.e., hardware manufacturers and hardware suppliers. The hardware supply chain threat actor has the ability to introduce vulnerabilities or malicious elements during the O-RAN manufacturing process or through the supply chain, such as creating backdoors or replacing benign hardware components with malicious ones. This risk is significant due to the potential for backdoor installations or unauthorised modifications to hardware components, firmware and operating systems, which are often deeply embedded in hardware or software layers that are difficult to detect and prone to espionage, data leakage and destructive attacks. The challenge for such threat actors is to guard against these risks in environments where it is inherently difficult to distinguish between benign and harmful components.

    \item \textbf{Independent Software Vendor (ISV)}: ISV plays an important role in the O-RAN ecosystem by providing infrastructure and application software. In O-RAN, ISVs are responsible for developing and delivering the various software components necessary for network operation, mainly including operating systems, orchestration frameworks, runtime environments, and specific O-RAN system components. ISVs contribute to the cloud-enabling and openness of the O-RAN architecture, enabling scalability and supporting new functionality needed for evolving use cases. ISV threat actors include deploying malicious logic in software or intentionally or unintentionally creating vulnerabilities that can be exploited. This covers a wide range of software elements in the O-RAN from operating systems to third-party applications, making it a key area of potential threat due to its complexity and widespread access within the network.

    \item \textbf{User Equipment (UE)}: UEs in O-RAN environments have traditionally been viewed as personal cellular phones, but have changed significantly in 5G and emerging 6G environments. With the advent of the Internet of Things (IoT), UEs now encompass a wide variety of devices (e.g., drones, smartwatches, security cameras, etc.), greatly expanding the attack surface of cellular networks. In O-RAN, UEs interact with the network by consuming RAN services, and their increased variety brings new security considerations. The diversity of UE types and the increased connectivity potential raise the risk of attacks, including the risk of cyber attacks such as DoS attacks by threat actors using UEs.

    \item \textbf{External Threat}: This aggregates a number of different unique types of threats that do not fall into other more clearly defined categories of threat actors. Common external threats can be categorised as: state-sponsored actors engaging in espionage or sabotage for strategic gain; corporations engaging in industrial espionage for competitive advantage; hacktivists targeting telecommunications to make a political statement; and terrorist organisations aiming to disrupt or disseminate information. These external threats can severely impact the O-RAN by undermining network integrity, stealing sensitive information, disrupting services, and exploiting vulnerabilities for harmful purposes. The diversity of motivations and resources available to these.

    \item \textbf{Insider Threat}: The ``insider'' category in O-RAN refers to privileged system-level users, such as employees or contractors who have programmatic access to O-RAN systems. Corrupt insiders can pose a significant risk because they may have broad access rights that allow them to manipulate systems, intercept or alter sensitive data, or even disrupt network operations. The breadth of these permissions and access rights makes it critical to implement strict security protocols and continuous monitoring to mitigate the risks posed by potential insider threats.

\end{itemize}

\subsubsection{Security Comparison Between 5G Traditional Interfaces and O-RAN Open Interfaces}

From the perspective of security architecture evolution, O-RAN is not a new system built from scratch, but rather constructed upon the existing interface definitions and security mechanisms of 3GPP 5G system. 3GPP SA3 established a layered security architecture based on TS 33.501 \cite{3gpp.33.501}, implementing unified authentication, encryption, and integrity protection mechanisms for key control interfaces including N1 (UE-AMF), N2 (gNB-AMF), and N11 (AMF-SMF). It implements unified deployment of authentication, encryption, and integrity protection mechanisms, achieving fine-grained key derivation and isolation through key systems such as K\_SEAF/K\_AMF. Specifically, N1 interfaces implement two-way authentication between UE and network via 5G-AKA or EAP-5G protocols, while encrypting and protecting the integrity of NAS signalling. N2, N11 and other interfaces are located in the controlled core network domain of operators, employing TLS, IPsec and Network Domain Security (NDS/IP) protocols to protect service-based signalling and user plane traffic. Mohammed et al. demonstrated that despite these mechanisms, these interfaces remain vulnerable to threats including signalling forgery and replay attacks, control plane DoS, routing tampering due to configuration errors, and leakage of sensitive identifiers and location information \cite{mahyoub2024security}. These risks remain typical security issues from 4G, presented in more complex interface types in 5G environments.

Building upon the above 5G security baseline, O-RAN introduces a series of open interfaces including the front-haul, A1, E2, O1, and O2, which make explicit the control and management interactions previously confined within single-vendor equipment. According to O-RAN Alliance WG11 security specifications \cite{oranSpec2024}, these interfaces continue to employ traditional cryptographic solutions at the underlying mechanism level: Fronthaul and F1 links are recommended to employ IPsec or TLS for encryption and integrity protection; control management interfaces such as A1, E2, O1, and O2 generally require TLS 1.2/1.3 mutual authentication based on PKI and X.509 certificate systems, supplemented by OAuth 2.0 and RBAC for authorisation control. O-Cloud and SMO enhance platform security through multi-tenant isolation, image signature verification, and secure boot mechanisms. From the perspective of the `control plane/management plane traffic encryption mechanism', O-RAN shares a high degree of origin with security mechanisms in traditional 3GPP interfaces. Therefore, threats such as signal flooding and forged control messages are essentially risks inherited from the traditional RAN and 5G core network interface, merely migrated to new open interfaces. Both the BSI's 5G RAN risk analysis \cite{BSI5GRAN2023} and the NTIA's Open RAN security report \cite{NTIAOpenRAN2023} indicate that many issues frequently attributed to new Open RAN risks-such as software vulnerabilities and supply chain threats-are equally present in closed RAN architectures. The distinction lies only in the manner of exposure.

The convergence of multi-vendor, virtualisation and open interfaces in O-RAN has altered trust boundaries, reshaping threat landscapes. O-RAN decomposes control and management functions into a pluggable xApp/rApp ecosystem and multi-tenant O-Cloud architecture, introducing more complex attack vectors. For instance, attackers may leverage malicious images, compromised third-party libraries, or CI/CD toolchains to launch supply chain attacks. Furthermore, the multi-tenant environment of O-Cloud and its open-source virtualisation stack introduce additional risks of side-channel attacks and cross-slice data leakage. Both the NTIA's \cite{NTIAOpenRAN2023} and O-RAN Alliance's \cite{oranSpec2024} reports emphasise that certain security controls within the specifications are defined as optional (such as the non-mandatory deployment of IPsec/TLS on specific interfaces and variations in the implementation of authentication strength). In multi-vendor collaboration scenarios, this constitutes a new architectural risk at the specification level - which is a dimension rarely covered in traditional 5G interface security analyses.

Table \ref{tab:interface_threats_comparison} provides a systematic comparison between traditional 5G interfaces and O-RAN open interfaces, clearly distinguishing between inherited risks and novel architectural risks.

\begin{table*}[t]
\centering
\caption{Security Baselines and Threat Landscape: Traditional 5G vs. O-RAN Interfaces}
\label{tab:interface_threats_comparison}

\scriptsize
\renewcommand{\arraystretch}{1.3}

\newcolumntype{Y}{>{\raggedright\arraybackslash}X}

\begin{tabularx}{\textwidth}{@{} p{1.6cm} p{2.2cm} Y Y Y @{}}
\toprule
\textbf{Interface} & 
\textbf{Endpoints} & 
\textbf{Security baseline (typical)} & 
\textbf{Inherited threats \newline (4G/5G-like)} & 
\textbf{O-RAN-specific threats \newline (openness/disagg./multi-vendor)} \\
\midrule

5G N1 & 
UE / AMF & 
5G-AKA / EAP-5G; NAS int+ciph & 
Signaling abuse/flooding; replay/downgrade attempts; ID/location leakage & 
N/A \\
\addlinespace

5G N2 & 
gNB / AMF & 
NDS/IP (e.g., IPsec); node auth; CP msg integrity & 
CP DoS; rogue/unauth gNB attempts; misconfig exposure & 
N/A \\
\addlinespace

5G N11 & 
AMF / SMF & 
SBA TLS/mTLS; PKI; API access control & 
API abuse; credential theft; request flooding & 
N/A \\
\addlinespace

\midrule 

O-RAN Open FH & 
O-RU / O-DU & 
Link protection (IPsec/MACsec where feasible); device auth; time-sync integrity checks & 
Rogue RU/DU; FH DoS; config tampering & 
Time-sync manipulation (PTP/SyncE); weakest-link defaults across vendors; interop gaps \\
\addlinespace

O-RAN A1 & 
non-RT RIC / near-RT RIC & 
mTLS; PKI; authZ (RBAC/OAuth2); policy integrity logging & 
MITM/replay (if misconfig); API DoS & 
Policy injection/poisoning; rApp supply-chain risks; cross-vendor policy conflicts \\
\addlinespace

O-RAN E2 & 
near-RT RIC / O-CU / O-DU & 
Secure transport (mTLS/IPsec where deployed); xApp identity+authZ; audit logs & 
Msg manipulation/replay; DoS; config errors & 
Malicious/compromised xApps; model-driven control abuse; CI/CD artifact poisoning \\
\addlinespace

O-RAN O1 & 
SMO / NMs & 
TLS for mgmt protocols; strong authN+authZ; secure logging/auditing & 
Unauthorized access; credential abuse; telemetry/log exposure & 
Vendor extensions \& semantic mismatch; baseline hardening misalignment; weakest-link ops \\
\addlinespace

O-RAN O2 & 
SMO / O-Cloud & 
mTLS + OAuth2/OIDC; multi-tenant isolation; image signing + SBOM gates & 
API abuse; privilege escalation; orchestration misconfig & 
Multi-tenant leakage/escape; container supply-chain/CVE blast radius; orchestrator compromise \\

\bottomrule
\end{tabularx}
\raggedright
{\footnotesize\emph{Abbreviation: } UE: user equipment; gNB: 5G base station; AMF/SMF: core control functions; FH: fronthaul;
mTLS: mutual TLS; PKI: public key infrastructure; RBAC: role-based access control; authN/authZ: authentication/authorization;
OIDC: OpenID Connect; SBOM: software bill of materials; CP: control plane; DoS: denial of service; NMs: Network Management Systems.}
\end{table*}

\subsubsection{Classification Method of Security Threats Related to O-RAN}
\begin{figure}[t]
\centering
\label{stride}
\includegraphics[width=3.5in]{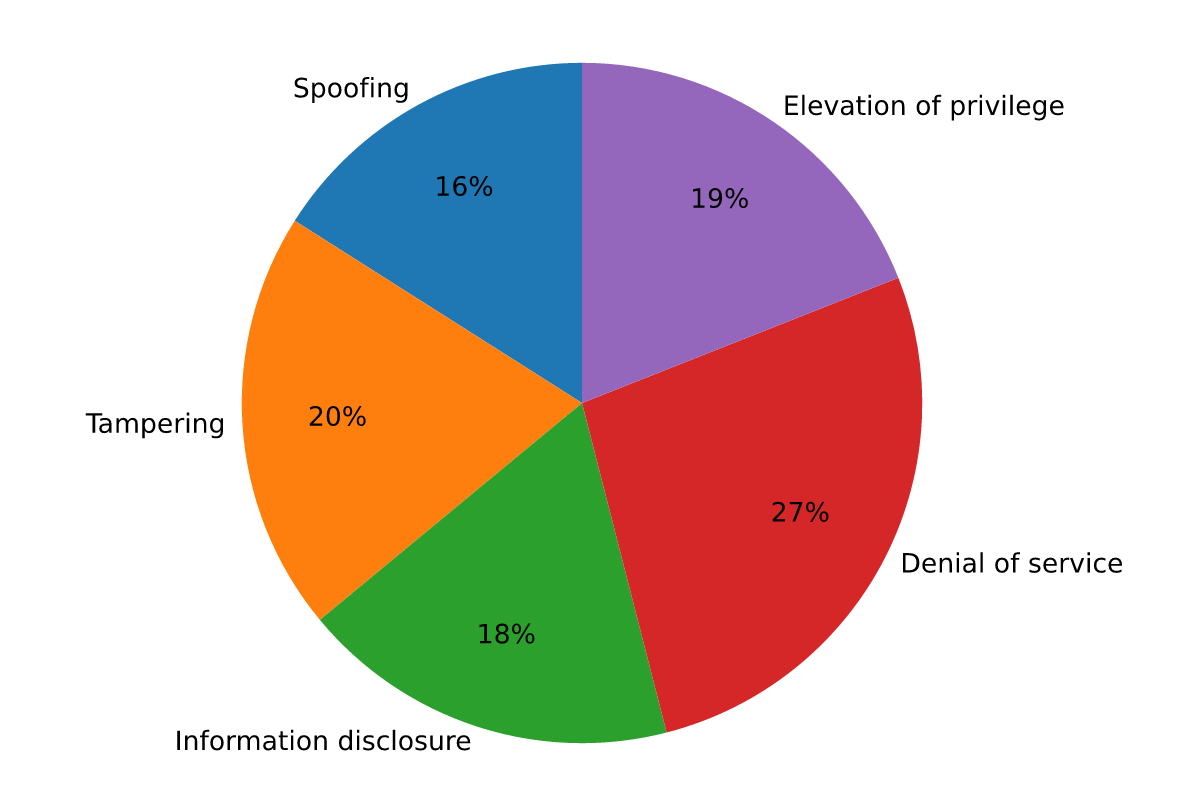}
\caption{Threat distribution in O-RAN based on STRIDE classification \cite{NTIAOpenRAN2023}}
\end{figure}

Following the previous comprehensive comparison of O-RAN security analyses published by international organisations and institutions, this subsection will delve further into the specific security threats existing in the O-RAN. In order to identify and analyse these threats more effectively, this subsection will adopt the STRIDE model and classify them according to the main functional areas of the O-RAN. This approach aims to break down complex O-RAN security issues into more manageable parts, thus providing more targeted and practically applicable security analyses. Fig. \ref{stride} depicts the distribution of threats in O-RAN based on the STRIDE classification, as statistically analysed by NTIA's O-RAN Security Report.

The STRIDE model \cite{Kohnfelder1999TheProducts}, a widely recognised security threat classification methodology, provides a systematic way to identify and articulate security vulnerabilities. The model includes six categories: Spoofing, Tampering, Repudiation, Information Disclosure, Denial of Service, and Elevation of Privilege. By using STRIDE, readers are able to more clearly understand and describe the various security threats that may arise in O-RAN, thus laying the foundation for developing effective defence strategies. Table \ref{securitythreats} provides a detailed description of the STRIDE model in O-RAN.

\begin{table*}[t]
\centering
\caption{STRIDE Security Threats in O-RAN}
\label{securitythreats}

\small 
\setlength{\tabcolsep}{6pt}
\renewcommand{\arraystretch}{1.25}

\newcolumntype{Y}{>{\raggedright\arraybackslash}X}

\begin{tabularx}{\textwidth}{@{} l Y Y @{}}
\toprule
\textbf{Threat Type} & \textbf{O-RAN-Oriented Definition} & \textbf{O-RAN Typical Manifestations} \\
\midrule

Spoofing &
An attacker impersonates a legitimate network entity, application, or management endpoint to gain access or inject control actions. Open interfaces and multi-domain identity chains in O-RAN make endpoint impersonation more prominent. &
Impersonating an xApp or rApp to obtain control capabilities; spoofing a peer service or node on the A1 or E2 interface; impersonating a management identity to access SMO or the O-Cloud control plane. \\
\addlinespace

Tampering &
Unauthorized modification of policies, configurations, software artifacts, or telemetry leads to control deviation, incorrect decisions, or abnormal system behavior. Cloud-native delivery and the application ecosystem in O-RAN make software and configuration tampering a key risk source. &
Tampering with A1 policies or SMO configurations to alter control logic; tampering with xApp or rApp images and dependencies to introduce malicious behavior; tampering with telemetry or control messages to degrade closed-loop effectiveness. \\
\addlinespace

Repudiation &
A principal denies performing an action when verifiable audit evidence is insufficient, resulting in weak attribution and accountability. Multi-vendor and cross-domain orchestration complicate consistent auditing in O-RAN. &
Policy issuance and configuration changes lack verifiable audit trails; application deployment and rollback lack a consistent chain of responsibility; distributed logs hinder cross-domain correlation and forensics. \\
\addlinespace

Information Disclosure &
Sensitive data, critical configurations, or security credentials are accessed or exposed without authorization. O-RAN expands the surface for telemetry, policy, and management data flows, increasing the impact of leakage. &
Exposure of user data, location data, or operational telemetry; exposure of policies, model outputs, or training-related data; leakage of certificates, keys, or access tokens enabling further lateral movement. \\
\addlinespace

Denial of Service (DoS) &
An attacker exhausts resources or disrupts critical paths to make control or management functions unavailable. O-RAN expands target surfaces through cloud resource pools and open control and management interfaces. &
Resource exhaustion against the O-Cloud causing service unavailability; flooding attacks against SMO or the RIC platform disrupting orchestration and control; request flooding on key interfaces causing performance degradation. \\
\addlinespace

Elevation of Privilege &
An attacker escalates from low privilege to high privilege and gains platform-level control, impacting applications, policies, and infrastructure operations. Multi-tenancy and cloud-native authorization in O-RAN amplify the consequences of misconfigurations. &
RBAC misconfiguration enables cluster-level or platform-level privilege; abuse of token scope or authorization paths to obtain elevated management rights; lateral movement from the application runtime to orchestration and infrastructure layers. \\

\bottomrule
\end{tabularx}
\end{table*}

\subsubsection{Security Risks Related with O-RAN}

Spoofing attacks pose a serious security threat in O-RAN architectures because they can manipulate or falsify the identities of network entities, thereby affecting the operation and security of the entire network. Specifically, Spoofing attacks can affect O-RAN at multiple levels such as network control, data security, etc. At the network control level, an attacker through Spoofing may forge control signals or device identities, resulting in resource allocation errors, signal interference, and even service interruption, especially in the CU, DU, and RU in these core components. Moreover, in terms of data security and privacy, Spoofing attacks may pose a threat to critical components such as O-Cloud and SMO. By masquerading as a legitimate node, attackers are able to access or tamper with sensitive data and network configurations, thereby jeopardising data confidentiality and integrity. Common security threats include pseudo-base station attacks \cite{Park2023SMDFbs:Stations}, signalling masquerades, Rogue O-RUs, RAN spoofing \cite{openran2021security}, and attacks on the master clock \cite{Dik2021TransportFronthaul}. In order to ensure that the communication and operation between the components and interfaces of the O-RAN are not affected by Spoofing, it can be done by strengthening the authentication and authorisation mechanisms, network monitoring, anomaly detection and other strategies.

Of the 1,338 unique security threats to O-RANs identified by the NTIA organisation, Tampering accounted for 20\% of the overall total, ranking second out of the six security threat types \cite{NTIAOpenRAN2023}. The impact of Tampering in O-RANs is primarily in the form of unauthorised alterations to hardware, software and data. Hardware tampering affects the physical components of the O-RAN, such as RUs, CUs and DUs. Such tampering may introduce backdoors or other security vulnerabilities through the addition or modification of hardware components, leading to communication disruption or signal hijacking, which in turn compromises the integrity of the physical equipment. Software tampering may involve the installation of malware (e.g., Malicious xApps and Malicious rApps), the modification of configuration files, or the alteration of system settings, which may result in data leakage, abnormal system operation, or degradation of network performance. Unauthorised changes to data transmitted in the O-RAN affect all components that depend on data transmission, including RUs, CUs, DUs, RICs, etc. Common vulnerabilities include UE identification due to malicious xApps \cite{alliance2021ran}\cite{ericsson2023openran}, data poisoning attacks \cite{Hu2022MembershipSurvey}, model poisoning attacks \cite{Shi2021AdversarialSlicing}\cite{Iturria-Rivera2022Multi-AgentO-RAN}, etc. Common security strategies include regular software integrity checks and data encryption, data validation, machine learning model monitoring, etc. With these comprehensive security strategies, the potential risks posed by tampering to O-RAN systems can be significantly mitigated to ensure the efficient and secure operation of the network.

Repudiation in O-RAN architectures mainly involves network participants (e.g., users, devices, or operators) denying their previous actions or transactions, especially when these actions result in undesirable consequences. The lack of effective logging and auditing mechanisms in various key components of the O-RAN may lead to ambiguous traceability of operations and attribution of responsibility. For example, the threat of repudiation is particularly pronounced in data- and management-intensive components such as O-Cloud and SMO, which involve a large number of data processing and network management operations. Untracked data access and modifications can result in the disclosure of sensitive information or compromised data integrity, while the accuracy of network management and control decisions can be compromised by denial or tampering with the operations involved. Common security threats against repudiation include log tampering (where an attacker deletes or modifies log records to hide his malicious activities), identity forgery and activity denial (where an attacker forges an identity to carry out an attack activity and then denies his actions), and the absence of audit trails (where critical operations such as configuration changes and network optimisation decisions cannot be traced) \cite{shahbazi2022analysis}. Therefore, in order to maintain the overall security of O-RAN, robust logging and auditing mechanisms need to be implemented, which not only ensures that all operations can be traced and verified, but also strengthens authentication and access control, and improves the transparency and traceability of network operations.

In O-RAN architecture, Information Disclosure involves sensitive data or critical information being accessed by unauthorised individuals or entities. Such disclosure can affect various components of an O-RAN, especially those that process and store large amounts of user data, network configuration, and security credentials. For example, CUs and DUs handle user data such as communication content and location information, the compromise of which could violate user privacy and breach data protection regulations \cite{regulation2018general}. In sections such as O-Cloud and SMO, critical network configuration and management information, if compromised, could lead to unstable network operations and large-scale leakage of user data, increasing legal and reputational risks \cite{Liyanage2023OpenOpportunities}. Similarly, the leakage of communication protocols and interface specifications in O-RAN's open interfaces may provide attack vectors for attackers, enabling them to launch attacks against the network more effectively \cite{openran2021security}. To defend against these threats, O-RANs need to take measures that include data encryption, strict access control, the fronthaul of O-RANs can follow the MACsec suite of protocols, and network and application layer network automation can be deployed with SDS-based approaches \cite{klement2022open}\cite{blanc2018towards}. These measures help reduce the risk of information leakage and protect the security of the entire network.

Denial of Service (DoS) attacks, especially in their distributed form (DDoS), pose a serious security threat to O-RAN architectures and are designed to make network services unavailable. These types of attacks are particularly dangerous because they have a direct impact on the stability and performance of the network. And with the development of 5G and 6G technologies, the rapid growth in the number of UEs may increase the risk of DoS attacks. In the open ran security report organised by NTIA, they identified a total of 1,338 unique security threats related to O-RANs, with the largest percentage of threats (27\%) related to denial of service \cite{NTIAOpenRAN2023}. Key components of O-RANs, such as CUs, DUs, RUs, and real-time and non-real-time RICs, could be the targets of DoS attacks \cite{Liyanage2023OpenOpportunities}. These attacks may cause these components to be unable to process signals or data properly, thus affecting the operation of the entire network. In particular, O-Cloud, as the data processing and storage centre of the O-RAN, is particularly sensitive to DoS attacks, which may lead to the exhaustion of its resources, making critical data processing and storage services unavailable. Similarly, SMO is responsible for network management and service orchestration for the entire O-RAN, and a DoS attack could seriously disrupt its operations. To counter these threats, O-RANs need to take a variety of measures including traffic monitoring, filtering and firewall protection, as well as implementing strategies to defend against DoS attacks in the network design, such as capacity planning and redundancy configurations.

Elevation of Privilege (EoP) in the O-RAN architecture involves unauthorised users or processes gaining a higher level of access than is normally allowed. The components of O-RAN that are most susceptible to the EoP threat include centralised data processing and management systems such as O-Cloud and SMO, where attackers can exploit infrastructure vulnerabilities due to poor maintenance or misconfigurations to look for any authorisation violations and gain higher levels of access \cite{Ranaweera2021SurveyPrivacy}. With this high-level access, attackers are able to go from simply over-allocating resources to completely removing xApps or rApps \cite{alnaim2019misuse}\cite{Qiang2018PrivGuard:Attacks}.  Attackers may also change the configuration of a compromised VNF/CNF to consume large amounts of CPU, hard drive, and memory resources to exhaust the hypervisor. Or the attacker generates excessive log entries in the VNF/CNF to make log file analysis more difficult. Finally, when attackers have root access to the hypervisor, they can extract user identities (IDs), passwords, and Secure Shell Protocol (SSH) keys from memory dumps by using search operations, which in turn violates user privacy and data confidentiality \cite{Yang2016AVirtualization}\cite{Khan2020ADirections}. To address these threats, implementing strict access control policies and strong security protocols is critical to maintaining the overall integrity and security of the O-RAN network. Regular security assessments and updates are also critical to address potential vulnerabilities that could be used to elevate privileges.

\subsection{Privacy Challenges}
\subsubsection{Personal Data Exposure Risks}
O-RAN has significantly improved the flexibility and efficiency of wireless networks through open interfaces and standardised protocols, but the openness of O-RAN has also allowed multiple providers and entities to access network data, which has led to ensuring user privacy becoming challenging. And with the rise of new technologies, architectures, and services in 5G and 6G, there will be more user devices as well as greater volumes of data, which will undoubtedly make privacy protection more difficult. This includes not only the protection of user data, but also the security of identity and personal information \cite{Sorensen20155GPrivacy}. Users' privacy-sensitive data is usually leaked through communication services that collect various types of personal information, which is often not necessary for the operation of the services. In addition, attackers may further extract more personal information about the user, such as UE application lists, location information, and search/shopping preferences \cite{Liyanage2023OpenOpportunities}. The multi-vendor environment of O-RAN further increases the complexity of privacy protection, where different entities may have different privacy protection standards and practices, which requires a unified privacy protection framework to ensure data security across the network. Thus, the importance of privacy in O-RAN stems not only from technical and operational needs, but also from the necessity to comply with legal requirements and maintain user trust. Ensuring privacy will help O-RAN achieve its goal of comprehensive network innovation while maintaining its commitment to the security and integrity of user data.

Privacy risks not only come from direct access to individual user data, but are further reflected in the challenges of data governance across vendors and domains. To support intelligent decision-making for Near-RT RIC and Non-RT RIC, operators need to continuously collect and share fine-grained RAN telemetry and KPIs (such as cell-level load, UE-level latency and packet loss patterns) between different xApps/rApps. Some research and industry solutions have proposed establishing RAN data lakes across multiple domains and operators to collect RAN telemetry and KPIs from diverse networks and vendors \cite{carrozzo2020ai} \cite{5g2017view}. This facilitates cross-domain AI/ML model training and joint optimisation. In such environments, simple rely on interface encryption or access controls is not enough. Fine-grained data governance rules must be established, including who is the data controller and who is the data processor. Otherwise, RIC ecosystem will remain in a state of weak governance.

Furthermore, research indicates that even aggregated or pseudonymised mobility traces can be re-identified for the majority of users with only a small number of location points when the spatio-temporal resolution is high enough \cite{de2013unique}. This means that RIC datasets constructed based on cell IDs, latency distributions, mobility patterns, and other factors are difficult to render fully anonymous. When such data is further centralised within O-Cloud or cross-regional data centres, regulations such as GDPR may classify part of it as personal data, thereby triggering stringent requirements for data minimisation, purpose limitation, and cross-border transfers. In multi-vendor, cross-cloud O-RAN deployments, the challenge of implementing regulatory requirements such as data localisation and cross-border transmission clauses while meeting performance and intelligent optimisation demands has given rise to a new category of privacy challenges. This will be explored further in Chapter VI.

\subsubsection{AI-Related Privacy Issues}
With the continuous development of AI technology breakthroughs in recent years, AI has become a powerful technology and plays an important role in O-RAN, mainly for optimising network performance and management, such as improving network efficiency and reliability through intelligent traffic management, predictive maintenance, resource optimisation, and security management \cite{lin2023embracing}. However, AI may also pose privacy risks as they typically need to handle large amounts of personal and sensitive data. Unexplained AI systems may inadvertently leak user data if used in O-RANs, or lead to unintended consequences if certain data is incorrectly used in an opaque decision-making process \cite{Yampolskiy2020UnexplainabilityAI}\cite{Hamon2022BridgingDecision-Making}. In addition, automated decision-making by AI may be exploited maliciously, e.g., by tampering with xApps or rApps to manipulate data or functionality (e.g., UE identification due to malicious xApps \cite{Liyanage2023OpenOpportunities}), which also increases the risk of misuse or leakage of user data. Thus, while AI offers significant advantages in O-RAN, the privacy protection issues associated with it also need to be handled carefully.

In practice, the capabilities of such AI systems often rely on long-term collection of large-scale RAN telemetry and user behavioural characteristics, which are used for centralised training and continuous retraining of xApp/rApp models on O-Cloud or RIC. Due to the absence of data specifications towards O-RAN, operators and vendors face difficulties in defining which characteristics may be retained long-term for model training, and which may only be used for online inference and fault diagnosis. Existing standards also rarely address the protection requirements against AI-specific privacy attacks, such as model back-inference, in RAN scenarios. These practical constraints make achieving verifiable privacy protection while maintaining model performance a key unresolved challenge in current O-RAN systems.

\subsubsection{O-Cloud Privacy Challenges in Multi-Tenant Environments}
O-cloud component plays a critical role in the O-RAN architecture by applying the principles of cloud computing to the wireless network environment. O-cloud provides a physical computing platform for creating and hosting Virtual Network Functions (VNFs) and Cloud Network Functions (CNFs), which support the key elements of O-RAN, such as Near-RT RIC, O-CU-CP, O-CU-UP, and O-DU \cite{Polese2023UnderstandingChallenges}. However, it shows that the privacy protection challenges of Cloud Computing are also relevant to Open RAN. Firstly, the storage and processing of user data is no longer confined to the local system but shifted to the cloud service provider's servers. which may result in the transfer of data control from the user and increase the risk of data leakage or unauthorised access. In addition, since cloud services often span multiple geographic and legal jurisdictions, cloud service providers must ensure that their services comply with various data protection laws and regulations. Finally, the multi-tenant nature of cloud computing environments requires effective data segregation measures to prevent data leakage or cross-access between different customers \cite{6238281}\cite{6187862}.

\section{Emerging Challenges for O-RAN in 6G}
This chapter explored key security and privacy challenges encountered when deploying O-RAN within 6G environments. Based on 6G IMT-2030 kPIs listed in Table \ref{tab:5g_vs_6g}, we focused on how these performance targets reshape design constraints for O-RAN security mechanisms-such as latency budgets, resource overhead, and trust boundaries. Unlike Chapter III, which primarily analysed O-RAN threats within 5G architectures, this chapter highlights challenges that emerge or become critical only when O-RAN is deployed as part of 6G-era architectures (such as SAGIN, THz, ISAC, and AI).

\subsection{Network Heterogeneity and Complexity}
\subsubsection{Heterogeneous Network Management of O-RAN}

\begin{figure}[h]
\centering
\includegraphics[width=3.5in]{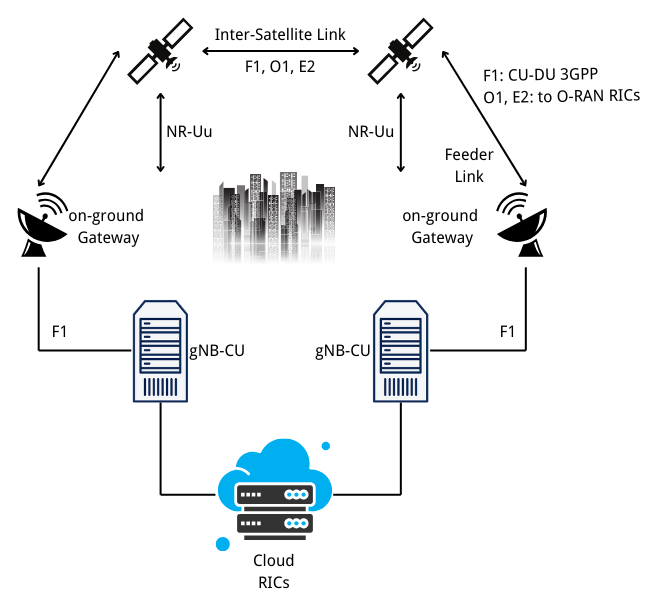}
\caption{O-RAN-based NTN system architecture}
\label{NTN}
\end{figure}

In the evolving field of wireless communication, there is a need to accommodate not only the communication needs of dense urban centers, but also to extend seamless connectivity to remote and underserved areas. To design such a network architecture means that different types of network technologies-such as cellular, Wi-Fi, and potentially non-terrestrial networks (NTNs)-will be integrated within a unified framework. Each network type has its own unique operating protocols, performance characteristics, and service capabilities. Such integration requires unprecedented network heterogeneity and complexity, demonstrating not only technological diversity but also reflecting the wide range of application scenarios, thus posing new challenges in design and management. There are currently studies underway to integrate non-terrestrial networks (NTNs) with terrestrial networks within the O-RAN framework \cite{rihan2023ran}\cite{choi2023spectrum}. NTNs, including satellite links, High Altitude Platform Stations (HAPS) and other aerial vehicles, play a crucial role in enabling global coverage, providing significant benefits such as extended coverage, enhanced capacity and resilience to terrestrial infrastructure failures \cite{Araniti2022TowardNetworks}\cite{Giordani2021Non-TerrestrialOpportunities}. This integration approach leverages the open interfaces and standardized design of O-RAN to not only extend network coverage, especially to reach remote areas, but also enables efficient management and optimization across terrestrial and non-terrestrial network resources by leveraging AI and ML techniques.

However, compared to traditional terrestrial cellular networks, the introduction of NTNs into the 6G O-RAN architecture imposes more challenging constraints on latency and reliability. Typical low Earth orbit (LEO) satellites operate at altitudes of approximately 200-2000 km, with one-way propagation delays typically ranging from several ms to tens of ms. In contrast, geostationary Earth orbit (GEO) satellites orbit at approximately 36,000 km, where one-way propagation delays often approach or exceed 250 ms, readily resulting in end-to-end round-trip delays surpassing 500 ms \cite{Giordani2021Non-TerrestrialOpportunities}\cite{Araniti2022TowardNetworks}. By contrast, IMT-2030 specifies a target latency range of 0.1-1 ms for 6G, requiring data transmission to be completed with a success probability at the $10^{-5} \text{--} 10^{-7}$ level under high-reliability conditions \cite{ITU-R_M.2160-0}. This means that once O-RAN control and data planes are extended to non-terrestrial nodes such as satellites and HAPS, the latency and computational budget available for security mechanisms will be significantly constrained. Meanwhile, link jitter caused by the orbital motion of LEO satellites further amplifies the risks of handshake timeouts, data retransmissions, and security policy failures. This represents a significant divergence between 6G-NTN scenarios and 5G terrestrial networks, directly impacting the feasibility of multiple existing protection mechanisms within 6G-O-RAN.

Campana et al. \cite{10188308} proposed an NTN (Non-Terrestrial Network) system architecture incorporating the O-RAN concept, as shown in Fig. \ref{NTN}. This architecture integrates the terrestrial, access, and user segments. The terrestrial segment interconnects ground network elements, such as gNBs, the 5G core, and globally distributed gateways, bridging the terrestrial network with the NTN segment. The access segment provides coverage through regenerative NTN nodes, with satellites carrying both RU and DU components. These satellites utilize either fixed or mobile beams and may host the complete or partial gNB functionality. This setup enables the deployment of O-RAN in space, with open interfaces like F1/E1 adapted for satellite connections and supported by feeder links. The RIC (RAN Intelligent Controller) components are cloud-based and are connected to the terrestrial network through the distributed network, allowing integration with ground elements. This architecture demonstrates the feasibility of applying O-RAN principles within satellite networks, facilitating more flexible and scalable NTN deployments.

From a security perspective, the integration of NTNs into 6G O-RAN not only inherits long-term security risks inherent in traditional satellite and aeronautical communication systems but also introduces unique architectural vulnerabilities specific to the 6G-O-RAN framework. On the one hand, the introduction of NTNs expands network attack surfaces by increasing new access points, forwarding links, and software/hardware supply chain segments, while exposing them to more potential attackers through complex cross-border transmission paths. On the other hand, O-RAN needs to schedule and manage highly heterogeneous terminals and nodes-such as ground base stations, HAPS, and UAV platforms-within a unified framework. These devices exhibit significant differences in physical protection capabilities, computational resources, and security requirements. If such heterogeneity is disregarded and only security baselines designed for traditional ground terminals are applied, any compromise of a single weak link could evolve into a systemic risk affecting the entire SAGIN end-to-end communication.

Research indicates that traditional satellite and aeronautical communication systems face certain security threats that may compromise system confidentiality and integrity \cite{maurer2022security}. A typical example is Controller-Pilot Data Link Communications (CPDLC) \cite{alhalabi2025secure}: substantial volumes of messages are transmitted in plaintext over open channels, lacking robust authentication and access controls. It has been demonstrated that these communications can be forged or tampered with, thereby posing a direct threat to the authenticity and integrity of flight instructions. If vulnerable aeronautical/satellite links are directly connected to O-RAN's open interfaces, they will become vectors for attackers to compromise 6G O-RAN.

Besides historical legacy issues, incorporating NTNs as part of O-RAN introduces a series of 6G-specific architectural risks. First, when O-RU/O-DU units are deployed on satellites or HAPS, their physical security and supply chain integrity become more challenging to guarantee. Once a node is captured, cloned, or maliciously maintained, attackers may simultaneously implant backdoors at both the physical layer and protocol stack \cite{kumar2024review}\cite{abdelsalam2025physical}. Second, NTNs may involve cross-border data flows spanning multiple countries and regions, with data frequently relayed across different jurisdictions. This makes data sovereignty, log auditing, and liability attribution more complex \cite{abdel2022security}. Consequently, this means that once a privacy breach occurs, its scope of impact and scale of losses may exceed those of traditional terrestrial networks. The introduction of NTNs within O-RAN requires a redefinition of trust boundaries and threat models to ensure that while enhancing communication efficiency and coverage capabilities, overall security and privacy protections are simultaneously strengthened.

\subsubsection{Dynamic Resource Allocation}

Dynamic resource allocation is one of the core mechanisms within cellular networks. By adjusting the allocation of resources such as spectrum, power, and computing in real time according to service demands and changes in wireless conditions, it enhances overall resource utilisation efficiency and service quality. In O-RAN architecture, dynamic resource allocation is the key to network flexibility and on-demand customisation capabilities, enabling operators to achieve fine-grained scheduling on open, decoupled software and hardware platforms. Towards 6G, dynamic resource allocation must not only fulfil differentiated service level specifications in environments where multiple services and slices coexist, but also operate in scenarios such as URLLC/HRLLC \cite{Wang2023Resource6G}. This means that these mechanisms must be designed with sufficiently stringent latency and reliability budgets from the beginning, to prevent security detection, authentication, and reconfiguration processes from squeezing out latency margins for key business operations.

Unlike relatively slow manual configuration in traditional LTE/5G, dynamic resource allocation in O-RAN is typically driven by Near-RT RICs, whose control loops operate within a 10 ms to 1 s cycle. Policy updates are dispatched to O-DU/O-CU via E2 interfaces. xApp needs to complete model inference and resource decisions within this timescale, based on continuously refreshed key performance measurements (KPM) and user context. The actual physical layer scheduling remains performed by O-DUs within a $<10$~ms real-time loop \cite{Polese2023UnderstandingChallenges}. Therefore, any security mechanism coupled with dynamic resource allocation-such as real-time detection of anomalous traffic or malicious slice requests-must be accommodated within this stringent time and computational budget \cite{megarry2024understanding}. Too complex or frequent security operations would directly impact latency margins within near-real-time control loop, thereby conflicting with 6G's URLLC/HRLLC objectives.

The intelligent decision-making process of dynamic resource allocation ensures efficient utilization of key resources such as spectrum, computation and storage, while the network slicing technology further enhances the granularity and targeting of such resource management. In 6G networks, this combined use brings a number of advantages \cite{Wu2022AI-NativeNetworks}\cite{Zhang2019An5G}: first, by accurately controlling and allocating resources, the network is able to support a wider diversity of services, catering to different scenarios ranging from Enhanced Mobile Broadband (eMBB) to Ultra-Reliable Low-Latency Communications (URLLC) and massive Internet of Things (mIoT). Second, network slicing enables different services and applications to coexist on the same physical network infrastructure without interfering with each other, thus greatly improving spectrum utilization and overall network performance; finally, providing personalized resource allocation and network environments for users and services significantly optimizes the user experience and ensures that each service receives the network performance and quality of service guarantees it requires.

Several studies focus on AI/ML-driven optimization of resource allocation. Qazzaz et al. proposed a Random Forest Classifier-based ML model within the xApp framework, which is designed to optimize Physical Resource Block (PRB) allocation in response to traffic demand and QoS requirements, and to enhance the performance of the scheduler function within the O-DU \cite{qazzaz2024machine}. Mohammadreza et al. designed an xApp solution based on an Actor-Critic reinforcement learning model aimed at demonstrating the effectiveness of the rate in terms of near real-time control and automation in a production-like network setup \cite{10008713}. These studies reveal the potential of dynamic resource allocation techniques in O-RAN.

Although dynamic resource allocation and network slicing are anticipated to deliver flexibility and efficiency gains for O-RAN and 6G networks, their combination also expands the attack surface across both the control plane and data plane. In O-RAN, xApps on near-RT RICs usually collect substantial fine-grained telemetry data, slice requests, terminal locations, and other information to learn or infer resource allocation policies \cite{qazzaz2024machine}. Once such telemetry data or slicing requests are maliciously tampered with, it may trigger data/model poisoning attacks or policy deception. Research shows that by deliberately perturbing environmental observations of DRL-based resource allocation agents, PRB allocation and throughput for target services can be significantly reduced \cite{habler2025adversarial}\cite{ergu2024efficient}. 

Meanwhile, user context data processed during dynamic resource allocation contains substantial privacy-sensitive information, such as user identity and location details. In scenarios where multiple tenants share infrastructure, logical isolation between slices is typically implemented while underlying physical resources (such as processor caches) remain shared. Cross-slice side-channel attacks can infer sensitive information belonging to tenants of other slices by probing low-level characteristics such as cache hit rates and queue delays, thereby posing risks of cross-slice data leakage \cite{abood2023classification}. Li et al. conducted systematic analysis of such side-channel risks in 5G RAN scenarios, and proposed an SCA-aware resource allocation algorithm \cite{li2019side}. Wei et al. employed reinforcement learning to automatically search for optimal caching side-channel attack strategies \cite{shao2024attacking}.

These findings show that traditional network security policies employing static slice access control, coarse-grained rate limiting, or simple resource cap constraints offer limited protection against the dynamic, AI-driven resource allocation and multi-tenant virtualisation environments characteristic of 6G O-RAN.

\subsection{Advanced Technology Vulnerabilities}
As 6G networks evolve, O-RAN may integrate several key advanced technologies to meet network requirements. However, while the introduction of these technologies significantly improves network performance, it also inevitably brings new security vulnerabilities and threats. Some of the technologies that are likely to be integrated and the security challenges they pose include:
\begin{itemize}
    \item \textbf{Terahertz Communication: }Terahertz communication is one of the key candidate technologies for 6G, which works in the approximately 0.1-10 THz frequency band. Compared to millimetre waves, it offers wider usable bandwidth, supporting data rates of 100 Gbps or greater levels. Through narrow beams, it enables high spatial resolution transmission, promising to meet the demands of scenarios such as immersive communications and ultra-high-capacity backhaul \cite{abdel2022security}. In O-RAN architectures, terahertz links can be deployed for backhaul/fronthaul in busy cell scenarios, indoor hotspot coverage, and ultra-high-speed interconnections between data centres, thereby reducing spectrum pressure and enhancing overall throughput. However, the propagation characteristics of high-frequency signals also introduce new security threats. On the one hand, the wide bandwidth of terahertz band enables attackers to implement frequency-sweeping interference using broadband noise sources or multi-carrier signals, even without precisely aligning the carrier frequency, thereby reducing link reliability. On the other hand, while high directivity reduces the probability of passive eavesdropping at long distances, attackers can still achieve effective interception or even directional jamming within complex indoor multipath environments by entering primary reflection paths \cite{Ma2018SecurityLinks}. Meanwhile, terahertz links are very sensitive to obstructions and misalignment errors \cite{11028770}. Attackers can form covert DOS attacks through physical blocking, mirror reflections, or manipulating smart reflectors. Existing physical layer security schemes for 5G millimetre-wave communications typically assume narrow bandwidths, limited antenna scales, and far-field beamforming. Direct migration to terahertz bands presents challenges such as increased algorithmic complexity and near-field beam misalignment. This shows that 6G O-RAN must reassess the security boundaries of channel modelling, beam management, and anti-interference mechanisms when adopting terahertz communications.

    \item \textbf{Reconfigurable Smart Surface (RIS): }RIS controls the phase and amplitude of incident electromagnetic waves in real time by deploying programmable arrays composed of numerous passive elements across building surfaces \cite{pan2021reconfigurable}. It enhances link gain, coverage quality, and spectrum efficiency without introducing additional radio frequency chains. In 6G O-RAN, RIS can be utilised in combination with intelligent optimisation applications running on the RIC. By dynamically adjusting reflection factors based on network conditions, it can deliver beamforming gains of more than 10 dB in hotspot areas and enhance the user experience at the network edge \cite{pei2021ris}. However, the programmability and distributed deployment of RIS also brings new security and privacy risks. Once attackers gain control of RIS controllers through software vulnerabilities, they can maliciously redirect signal paths without increasing transmission power, constructing undetectable directional interference or eavesdropping channels \cite{khalid2025malicious}. RIS deployed in public spaces also face risks of physical damage and tampering, where attackers can hijack beams by replacing or obstructing individual elements. Existing 5G beamforming and relay security schemes often default to treating relays as trusted components, without considering RIS itself as part of the threat model \cite{mughal2025malris}. The lack of authentication and integrity protection mechanisms for RIS control channels is inadequate in the context of large-scale RIS deployments within 6G O-RAN. 

    \item \textbf{Integrated Sensing and Communication (ISAC): }ISAC integrates communication and sensing capabilities within a single waveform, spectrum, and hardware platform, enabling networks to deliver high-speed connectivity as well as high-precision environmental sensing. It achieves sub-metre or even centimetre-level positioning and high-resolution imaging in scenarios such as transportation and industrial control \cite{wang2022integrated}. In O-RAN architecture, ISAC-related functions are typically scheduled by RIC, thereby enhancing adaptive capabilities to the wireless environment while ensuring service QoS. ISAC's integrated sensing and communication characteristics also introduce more complex security and privacy risks to O-RAN. On the one hand, ISAC networks require continuous collection of high-dimensional perception data (such as detailed location trajectories and posture information). So even without traditional identification, it is sufficient for precise re-identification and behavioural profiling of users or vehicles \cite{qu2023privacy}. If the relevant data or features are misused, it will trigger privacy leakage and compliance issues. On the other hand, ISAC also introduces cross-domain attack surfaces. Attackers may manipulate, forge, or replay perception data within the shared waveforms of the communication-perception system to mislead the RIC's environmental assessment, thereby inducing incorrect resource allocation or security policies \cite{wu2024ai}. In safety-critical scenarios such as autonomous driving and industrial control, such crosstalk attacks may further evolve into real-world security incidents. Research shows that traditional security frameworks, which model communication and sensing systems separately, cannot cover such cross-domain attacks \cite{qu2023privacy}. ISACs require the introduction of dedicated security and privacy protection mechanisms (such as trusted measurement and differential privacy) during design to adapt to the communication-sensing cooperation scenarios of 6G-O-RAN.

    \item \textbf{Edge Intelligence: }Edge intelligence enables O-RAN to achieve highly autonomous resource scheduling and real-time optimisation decisions by decentralising AI inference and partial training capabilities to edge nodes near the terminal. While such distributed intelligence enhances network performance, it also expands the attack surface. Edge-side AI models and their training/inference data face multiple threats, including adversarial examples, data and model poisoning, model theft, and privacy leaks. Experiments have demonstrated the construction of malicious xApps on O-RAN testbeds, achieved by introducing minute perturbations to KPMs or spectrum maps stored within near-RT RIC databases \cite{chiejina2024system}. This reduces the accuracy of interference detection or resource allocation agents, leading to incorrect scheduling and performance degradation within the network. Research has also showed that by injecting carefully constructed data during the model training phase, DRL-based xApps can be induced to make control decisions that benefit attackers under specific trigger conditions \cite{sapavath2023experimental}. Moreover, edge nodes usually have limited computational power and energy budgets, making it difficult to support high-overhead multi-model redundancy or complex trusted execution environments. This makes it difficult to directly migrate many traditional cloud-based AI security hardening measures to the edge intelligence scenarios of 6G O-RAN. Therefore, when introducing edge intelligence, O-RAN requires new security designs in areas such as xApp/rApp lifecycle management and lightweight robust training. It cannot rely only on generic protections at the NFV layer from 5G.
\end{itemize}

Table~\ref{tab:6g_enablers_summary} provides a comparative summary of THz, RIS, ISAC and edge intelligence, illustrating the performance drivers, component touchpoints and additional security/privacy attack surfaces introduced by these four key 6G enabling technologies within O-RAN.

\begin{table*}[t]
\centering
\caption{Summary of advanced 6G enablers and new attack surfaces in O-RAN.}
\label{tab:6g_enablers_summary}

\small 

\renewcommand{\arraystretch}{1.3} 

\newcolumntype{Y}{>{\raggedright\arraybackslash}X}

\begin{tabularx}{\textwidth}{@{} p{1.5cm} Y Y Y Y Y @{}}
\toprule
\textbf{6G enabler} &
\textbf{Key KPI / performance driver} &
\textbf{O-RAN touchpoints} &
\textbf{New security attack surfaces} &
\textbf{Privacy impact} &
\textbf{Why 5G-era controls fall short} \\
\midrule

THz communications &
Ultra-wide bandwidth, ultra-high data rates, highly directional links &
THz fronthaul/backhaul; beam training and alignment &
Wideband/targeted jamming; blockage/alignment-induced DoS; reflection-path eavesdropping or interference &
Indirectly increases location/trajectory inferability &
mmWave assumptions do not hold for THz channels/hardware impairments; security/anti-jam overhead conflicts with tight latency budgets \\
\addlinespace 

RIS (reconfigurable intelligent surfaces) &
Coverage and link-gain improvement; spectrum/energy efficiency; edge throughput gain &
RIS controller and configuration channel; tight coupling with RIC/xApps &
Malicious reconfiguration enabling covert interference or assisted eavesdropping; physical tampering or unit replacement &
Fine-grained location inference and target tracking &
Conventional designs assume the reflector/relay is trusted; insufficient authentication, auditability, and verifiability for RIS configurations \\
\addlinespace

ISAC (integrated sensing and communication) &
High-accuracy positioning/sensing; improved autonomy via joint comms--perception &
Sensing data acquisition/fusion; RIC-coordinated comms--sensing scheduling; shared waveform/resources &
Tampered/replayed sensing cues mislead RIC decisions; tightly coupled comms--sensing failures; dual-purpose interference &
Sensing data enables re-identification and behavioral profiling &
Threat models that separate comms and sensing are incomplete; comms-only encryption cannot prevent inference/data-fusion leakage \\
\addlinespace

Edge intelligence &
Near-real-time closed-loop optimization; fast autonomous control &
near-RT RIC/xApps; KPM/telemetry pipelines; edge nodes in multi-tenant deployments &
Adversarial examples, poisoning, and backdoors; malicious xApp supply chain; model extraction &
Telemetry can reveal user/service profiles; cross-tenant inference risks &
Cloud-centric AI defenses are too heavy for edge latency/compute budgets; static policies cannot keep up with high-rate reconfiguration \\

\bottomrule
\end{tabularx}
\end{table*}

\subsection{Privacy-Preserving AI and Federated Learning}

In 6G O-RAN, RAN telemetry data (KPM/KPI and logs, etc.) alongside service context will be generated at finer granularity and higher frequency, distributed across multiple gNBs and multi-domain edge nodes. To meet data sovereignty and compliance requirements, more and more intelligent functions are adopting federated learning (FL) for local training, replacing raw data aggregation with model updates \cite{9060868}. And this can be combined with O-RAN's hierarchical control architecture to form a cascaded training loop between near-RT RIC and non-RT RIC/SMO. This enables the cross-site global model capability without centralised uploading of raw data. Existing research has demonstrated this `multi-E2 node local training-near-RT aggregation-non-RT global aggregation' solution \cite{rumesh2024federated}, validating its feasibility in KPI-based scenarios.

However, FL only changes the location of risk exposure, shifting the attack surface from centralised data lakes to the updated chain of `training-aggregation-distribution'. First, model/data poisoning and backdoor attacks are more invisible in FL. Attackers need only control a small number of clients to inject backdoors into the global model through methods such as model replacement, causing the model to output incorrect decisions under triggered conditions as desired by the attacker \cite{bagdasaryan2020backdoor}. Moreover, many privacy-preserving implementations rely on secure aggregation to conceal individual client updates. This conversely diminishes the server's visibility and auditing capabilities regarding anomalous updates, thereby increasing the difficulty of detecting model poisoning. And finally, traditional Byzantine-robust aggregation is not perfect. Research shows that effective local model poisoning can be constructed against these robust aggregation rules, causing the performance of the global model to degrade \cite{247652}.

Secondly, research has shown that training samples may be reconstructed solely from gradients or model updates, potentially even inferring whether a specific record or subject participated in training, thereby leaking sensitive information \cite{zhu2019deep}. A more practical challenge is that the larger scale of participants and greater heterogeneity of equipment and links in 6G O-RAN exacerbate convergence instability and robustness issues. Meanwhile, mechanisms employed to mitigate leakage and poisoning risks including secure aggregation, differential privacy, and robust aggregation often introduce additional communication/computational cost and accuracy loss \cite{almanifi2023communication}, which conflicts with the low-latency vision of 6G. Therefore, when adopting privacy-friendly AI in 6G O-RAN, trade-offs between privacy strength, robustness, and performance requirements must be considered in threat modelling and system design. Corresponding mitigation approaches are discussed further in Chapter V.

\section{O-RAN's Security Advantages and Security Strategies}
\subsection{Security Advantages of O-RAN}

The decoupled architecture and open interfaces of O-RAN do not mean more security, but they provide a set of foundations for security engineering that can be explicitly governed and verified:

\begin{itemize}
    \item Verifiability and auditability enabled by open interfaces and standardisation. Through clearly defined interface boundaries and security control requirements, O-RAN enables standardised implementation of capabilities such as authentication, encryption, integrity protection, replay protection, logging, and auditing around interfaces. It does not directly eliminate threats, but it can reduce the assessment blind spots caused by invisible interfaces and black-box implementations, thereby providing the conditions for locating, tracing, and addressing STRIDE-type threats.
    \item A1/A2 establish manageable policy and model entry points, supporting controlled modifications \cite{oranSpec2024}. A1 abstracts policy guidance and closed-loop coordination between Non-RT RIC and Near-RT RIC into a manageable interface. This enables the embedding of governance mechanisms such as version control, conflict detection, and rollback, thereby mitigating systemic risks arising from policy mismatches and unauthorised policy injection. A2 enables AI/ML models to possess auditable pathways throughout training, signing/verification, distribution, deployment, and runtime monitoring. Within multi-xApp/rApp scenarios, model governance transforms model updates from one-off operational activities into traceable lifecycle processes. This mitigates the risks of model poisoning, backdoor insertion, and model substitution propagating across multi-vendor ecosystems.
    \item Cloud-native architecture and slicing mechanisms provide means for isolation and resilience, though they rely on strict baseline. O-Cloud's virtualisation and multi-tenant isolation capabilities, coupled with the logical partitioning of network slices, offer a way to suppress lateral movement, enforce least privilege, and control failure domains. When an intrusion event occurs, these mechanisms can be employed to limit the scope of impact. However, existing reports also emphasise that different vendors may employ different configurations regarding interface encryption, certificate rotation, and cloud platform baselines \cite{NTIAOpenRAN2023}\cite{CISAOpenRAN2022}. This can result in the above-mentioned isolation and auditing capabilities being potentially negated by configuration gaps, creating a security gap between specifications and deployment.
\end{itemize}

Table~\ref{tab:oran_security_advantages} summarises the above security capabilities, along with the key prerequisites and considerations required for implementation.

\begin{table*}[t]
\centering
\caption{Security Advantages of O-RAN}
\label{tab:oran_security_advantages}
\small 
\renewcommand{\arraystretch}{1.3} 

\begin{tabularx}{\textwidth}{@{} p{3.5cm} >{\raggedright\arraybackslash}X >{\raggedright\arraybackslash}X @{}}
\toprule
\textbf{Core feature} & \textbf{Security implication} & \textbf{Preconditions / caveats} \\
\midrule

Open interfaces \& Standardization & 
Interface-scoped verification and auditing; clearer security responsibility boundaries & 
Requires uniform security profile across vendors; avoid ``optional control'' gaps \\
\addlinespace

A1 (policy loop) & 
Policy provenance, versioning, conflict checking, controlled rollout/rollback & 
Mutual authentication + signed policy + rate limiting + audit trail; prevent policy injection/misuse \\
\addlinespace

A2 (model lifecycle loop) & 
Model governance (integrity, provenance, staged deployment, post-deploy monitoring) for multi xApp/rApp & 
Model signing/verification + backdoor checks + least privilege; adds operational overhead \\
\addlinespace

O-Cloud (cloud-native) & 
Micro-segmentation, tenant isolation, faster patching and security automation hooks & 
Hardened cloud baseline (image integrity, runtime isolation, secrets); misconfig/supply-chain risks remain \\
\addlinespace

Network slicing & 
Logical isolation and blast-radius containment for services/tenants & 
Isolation is not automatic; shared dependencies and weak enforcement can break containment \\

\bottomrule
\end{tabularx}
\end{table*}

\subsection{Strategies to Enhance O-RAN Security}
\subsubsection{Baseline Hardening}
The decoupling and multi-vendor features of O-RAN increase the interface exposure surface and supply chain links. Meanwhile, configuration discrepancies often exist between O-RAN in specifications and actual deployments, resulting in system security being vulnerable to the weakest link. Therefore, before discussing security enhancement strategies, this review synthesises a set of deployable and verifiable baseline hardening requirements based on practical guidelines such as ETSI, NIST, and CIS. These requirements serve to constrain the minimum security standards for O-RAN systems during multi-vendor deployments.

From standards alignment perspective, O-RAN ALLIANCE WG11 specification ETSI TS 104 104 \cite{ETSI_TS_104_104} explicitly defines security requirements and corresponding security control sets categorised by O-RAN interfaces and components, serving as the anchor point for baseline controls. ETSI TS 104 105 \cite{ETSI_TS_104_105} provides security testing specifications to support verifiability and consistency assessments of baseline controls. While ETSI TS 104 107 \cite{ETSI_TS_104_107}, as the O-RAN security protocol specification, outlines protocol-level implementation requirements. Furthermore, baseline hardening should emphasise configuration consistency and compliance verification for security options, preventing security risks arising from vendors defaulting to disabling critical controls.

Baseline controls at the interface and identity levels are defined around the principle of the smallest available closed loop. For the management plane (O1/O2) and control plane (A1/E2, etc.), end-to-end two-way identity authentication and encryption shall be prioritised for implementation, supported by unified certificate and key lifecycle management. Within multi-tenant O-Cloud environments, security baselines should follow centralised identity management policies and the principle of least privilege. For instance, rApp/xApp permission boundaries should be concretised (e.g., callable APIs, accessible telemetry scopes) and enforced through access controls and granular audit logging.

Platform-level baseline controls are primarily implemented in O-Cloud. NIST SP 800-190 \cite{NIST_SP_800_190} provides systematic guidance on critical risks within the container ecosystem (such as images, registries, and orchestration layers) alongside mitigation recommendations, serving as a baseline reference for O-Cloud container security. Meanwhile, Kubernetes hardening guidance \cite{NSA_K8s_Hardening} from NSA/CISA emphasises key actions including least privilege, network isolation, strong authentication, and log auditing, which are applicable to cluster hardening practices in critical infrastructure environments.

At the software supply chain and DevSecOps level, NIST SP 800-218 (SSDF) \cite{NIST_SP_800_218} provides a practical framework of fundamental secure software development practices. This framework can be applied to govern the secure development, vulnerability management, and release governance of xApp/rApp and SMO peripheral components (e.g., threat modelling, dependency management). For the O-RAN multi-vendor ecosystem, the value of SSDF lies in formalising the chain of responsibility for `who fixes, how to verify, and how to deploy rollbacks' into processes and evidence, thereby mitigating the long-term risks of supply chain attacks and configuration drift.

Table \ref{tab:hardening_baseline} summarises the proposed baseline control domains, minimal control sets, and their scope of application across O-RAN components/interfaces.

\begin{table*}[t]
\centering
\caption{Baseline Hardening Domains and Minimum Control Sets}
\label{tab:hardening_baseline}
\small
\renewcommand{\arraystretch}{1.3} 
\begin{tabularx}{\textwidth}{@{} l >{\raggedright\arraybackslash}X >{\raggedright\arraybackslash}X l @{}}
\toprule
\textbf{Control Domain} & 
\textbf{Minimum Control Set (Recommended)} & 
\textbf{Primary Scope (Examples)} \\
\midrule

\textbf{Interface Protection} & 
Mutual authentication (mTLS); Transport encryption \& integrity; Anti-replay mechanisms; Unified certificate/key rotation \& revocation. & 
O1/O2, A1/E2, Critical Southbound Control Links \\
\addlinespace

\textbf{Identity \& Auth} & 
Least privilege principle; Fine-grained API permissions \& rate limiting; Strong auditing (Subject, Action, Object, Time, Result). & 
SMO, Non-RT / Near-RT RIC, xApp/rApp Lifecycle Management \\
\addlinespace

\textbf{Container \& Runtime} & 
Image scanning; Least privilege; Isolation \& network segmentation; Runtime monitoring \& logging. & 
O-Cloud (Containers, Host OS, Image Registry, Orchestration Layer) \\
\addlinespace

\textbf{Kubernetes Hardening} & 
Cluster authentication \& RBAC; Control plane/etcd hardening; Network policies; Audit logging; Baseline checks (CIS Benchmarks). & 
O-Cloud (Kubernetes Control Plane \& Workloads) \\
\addlinespace

\textbf{Supply Chain \& Release} & 
NIST SSDF practices; Build traceability; Dependency \& vulnerability closed-loop remediation; Release approval \& rollback strategies. & 
xApp/rApp, SMO Peripheral Services, CI/CD Pipelines \\
\addlinespace

\textbf{Verifiability} & 
Security test cases \& conformance assessment; Continuous compliance checks \& baseline drift detection. & 
Security Acceptance, Interoperability Testing, Continuous Ops \\

\bottomrule
\end{tabularx}
\end{table*}

To operationalize the baseline hardening domains in Table \ref{tab:hardening_baseline}, we further map them to an interface-by-interface checklist. Table \ref{tab:oran_interface_security} is derived from O-RAN WG11 \cite{oran_security_update_2025} security requirements/protocol profiles and is complemented with widely adopted cryptographic and logging baselines (NIST SP 800-52r2/800-77r1 \cite{NIST_SP_800_52r2}, RFC 8915 \cite{RFC8915}, IEEE 802.1AE \cite{IEEE_802_1AEdk_2023}, and NIST audit/logging guidance \cite{NIST_SP_800_53_r5}).

\begin{table*}[t]
\centering
\caption{Security Controls and Requirements for O-RAN Interfaces}
\label{tab:oran_interface_security}

\tiny 
\setlength{\tabcolsep}{2pt} 
\renewcommand{\arraystretch}{1.3} 

\newcolumntype{L}{>{\raggedright\arraybackslash}X}

\begin{tabularx}{\textwidth}{@{} 
    >{\hsize=0.6\hsize}L  
    >{\hsize=0.8\hsize}L  
    >{\hsize=1.0\hsize}L  
    >{\hsize=1.0\hsize}L  
    >{\hsize=1.0\hsize}L  
    >{\hsize=0.9\hsize}L  
    >{\hsize=0.9\hsize}L  
    >{\hsize=1.8\hsize}L  
@{}}
\toprule
\textbf{Interface} & 
\textbf{Threat classes} & 
\textbf{Authentication} & 
\textbf{Key type and rotation} & 
\textbf{Cipher suites profile} & 
\textbf{Time sync integrity} & 
\textbf{API rate limiting} & 
\textbf{Audit fields and recommended logging} \\
\midrule

A1 \newline (Non-RT RIC to Near-RT RIC) & 
Spoofing, tampering, repudiation, information disclosure, denial of service & 
Mutual authentication at transport layer, plus explicit authorization for policy actions & 
X.509 certificate based identity with automated rotation and revocation handling & 
Prefer TLS 1.3. If TLS 1.2 is required, restrict to modern AEAD suites and disable legacy negotiation & 
N/A & 
Per client quotas, burst control, and request size limits at the API entry point & 
Log principal identity, authorization decision, policy identifier and version, request fingerprint, outcome, and correlation identifier. Preserve immutable timestamps \\
\addlinespace

E2 \newline (Near-RT RIC to E2 termination or E2 node) & 
Spoofing, tampering, denial of service, repudiation & 
Mutual authentication for the E2 transport endpoint. Validate sender identity at E2 termination & 
Certificates preferred for endpoint identity. Rotate credentials with short validity and automate renewal & 
Use a hardened profile for the chosen transport protection. If IPsec is used, prefer ESP with AEAD and conservative lifetimes & 
N/A & 
Enforce subscription and message rate caps. Apply backpressure to prevent control loop collapse & 
Log peer identity, service model and message type, subscription identifiers, validation failures, replay detection counters, rate limit drops, and correlation identifier \\
\addlinespace

O1 \newline (SMO to managed functions) & 
Tampering, repudiation, information disclosure, denial of service & 
Strong operator and service authentication, role based access control for management operations & 
Certificates for service endpoints plus privileged account governance. Enforce rotation windows & 
Prefer TLS 1.3 for management channels. For legacy, enforce hardened TLS 1.2 configuration & 
N/A & 
Throttle configuration change frequency. Protect against management plane storms & 
Log actor identity, target object, action type, configuration diff hash, approval reference, outcome, and session identifier. Retain logs with integrity protection \\
\addlinespace

O2 \newline (SMO to O-Cloud) & 
Spoofing, tampering, information disclosure, denial of service, repudiation & 
Mutual authentication plus least privilege authorization for orchestration and lifecycle actions & 
Certificates for service identity. Rotate keys and restrict long lived credentials & 
Prefer TLS 1.3 for API channels. Disable weak negotiation and legacy ciphers & 
N/A & 
Apply admission control and per tenant request budgets. Prevent orchestration request floods & 
Log workload identity, image digest, signature verification result, requester identity, authorization decision, resource change summary, and correlation identifier \\
\addlinespace

Open Fronthaul M-plane \newline (O-RU to O-DU management) & 
Spoofing, tampering, information disclosure, denial of service & 
Link or transport mutual authentication appropriate for the deployment model & 
Use link layer keying if MACsec is deployed. Rotate session keys periodically & 
If MACsec is used, apply AES GCM based protection. If IPsec is used, apply ESP with AEAD & 
Monitor management plane timing dependencies. Ensure that management time sources are authenticated where relevant & 
Rate limit management transactions and reject malformed payloads early & 
Log RU and DU identity, link security state, management command identifiers, failures, and correlation identifier \\
\addlinespace

Open Fronthaul C-plane and U-plane \newline (control and user plane) & 
Tampering, denial of service, information disclosure & 
Authentication and integrity for the transport segment where feasible under latency constraints & 
Session keys with controlled lifetimes. Prefer automated key refresh & 
Choose cryptographic protection that matches latency budget. Avoid negotiation mechanisms that add jitter & 
See synchronization row below when PTP is carried on the same segment & 
Apply policing for bursts and malformed sequences. Enforce strict parsing & 
Log control plane anomalies, malformed sequences, replay indicators, and alarmable counters for stability analysis \\
\addlinespace

Open Fronthaul synchronization \newline (PTP timing distribution) & 
Spoofing, tampering, denial of service & 
Authenticate timing sources when possible. Constrain trusted set & 
N/A & 
N/A & 
Detect grandmaster changes, abnormal offsets, path delay jumps, and time step events. Use redundancy and holdover & 
Rate limit PTP management messages and drop unexpected timing profiles & 
Log grandmaster identifier, offset and delay metrics, state transitions, threshold crossings, and time alarms with stable timestamps \\

\bottomrule
\end{tabularx}
\end{table*}

\subsubsection{Blockchain-enabled O-RAN}
\begin{figure}
\centering
\includegraphics[width=3.5in]{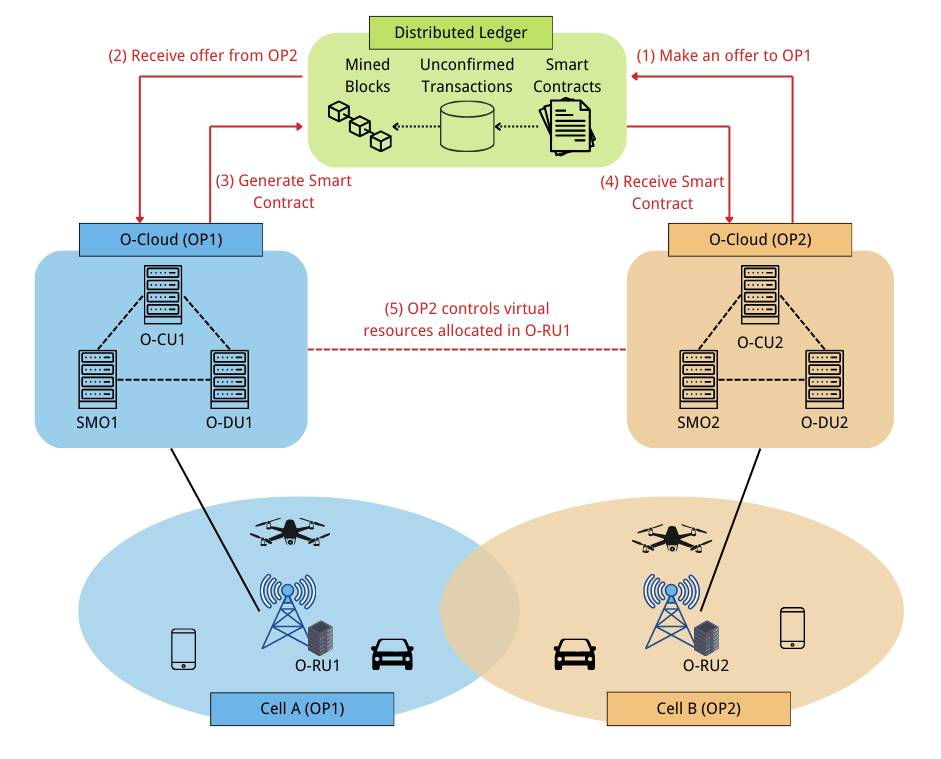}
\caption{A blockchain-enabled O-RAN sharing ecosystem}
\label{Blockchain}
\end{figure}

In O-RAN's multi-vendor ecosystem, blockchain is better suited as a trusted infrastructure for governance and evidence, supporting consistent record-keeping, traceable auditing, and accountability across participants. Fig. \ref{Blockchain} illustrates a blockchain-enabled O-RAN sharing ecosystem, a proposal based on the original 3GPP architecture for O-RAN, as introduced by Giupponi et al. \cite{Giupponi2022Blockchain-EnabledBeyond}. It consists of two operational clouds (O-Cloud OP1 and O-Cloud OP2), each associated with a different network operator and equipped with components such as the RIC, SMO, and O-CU. The use of distributed ledgers facilitates secure and transparent inter-operator transactions, utilizing smart contracts to dynamically automate service agreements and resource allocation. This setup enables operators to efficiently manage virtual resources across network segments (e.g., O-DUs and O-RUs), optimizing connectivity and quality of service for subscriber devices in various cellular coverage areas.

The value of blockchain for O-RAN primarily focuses on two scenarios. The first one is cross-operator RAN sharing and resource trading. Embedding the execution logic for resource leasing, settlement, and service level agreements into smart contracts reduces trust friction in cross-domain collaboration while enhancing auditability. Giupponi et al. propose a blockchain-enhanced O-RAN sharing architecture and discuss the benefits and trade-offs of embedding blockchain into O-RAN management processes \cite{Giupponi2022Blockchain-EnabledBeyond}. The second one is software supply chain and configuration change auditing. Fingerprint digests of critical objects such as xApps/rApps, CNF images, and policy versions are registered as tamper-proof records, providing a consistent basis for multi-party joint debugging and post-incident forensics. This approach aligns with industry requirements for O-RAN supply chain governance, such as O-RAN ALLIANCE's security updates explicitly mentioning vendor signatures and SBOM practices compliant with NTIA requirements \cite{oran_security_update_2025}.

At the privacy and compliance level, blockchain typically records only the minimum necessary metadata and hash summaries on-chain, with the original data stored in a controlled domain off-chain. Access authorisations and audit events are logged via smart contracts or policy engines. This on-chain/off-chain separation design enables participants to verify data integrity and prevent version rollbacks without sharing the original data. This provides a verifiable chain of responsibility and evidentiary foundation for cross-domain sharing. Furthermore, the consensus-based tamper-evident audit trail mitigates the risk of silent modifications to single-point logs in multi-party collaborations, thereby enhancing accountability and traceability within compliance frameworks. The multi-vendor collaboration and supply chain governance requirements of O-RAN further amplify the importance of evidence consistency and mutual trust in auditing.

Blockchain solutions also possess some limitations, with their practical boundaries primarily constrained by performance and security trade-offs. The first is latency and throughput constraints. Public chain PoW confirmation delays typically span minutes, whereas permissioned consensus mechanisms like PBFT and PoA can achieve sub-second confirmation under ideal conditions. However, this still cannot meet millisecond-level closed-loop control requirements, and thus should not be included in the control path of near-RT RICs \cite{jahid2023convergence}. The second is system overhead and scalability. Giupponi et al. note that maintaining a blockchain in a multi-party environment introduces computational, storage, energy consumption, and communication costs associated with block distribution, while also generating additional latency \cite{Giupponi2022Blockchain-EnabledBeyond}. These factors require explicit trade-offs during the design phase. Moreover, vulnerabilities in smart contracts, collusion among nodes, and majority control risks themselves constitute new attack surfaces, which require complementary contract audits, formal verification, and runtime monitoring \cite{chu2023survey}. Given these constraints, blockchain is better suited as a governance layer enhancement that complements baseline capabilities such as zero trust, interface hardening, and supply chain security. Its deployment must clearly define applicable interfaces and performance budgets.

\subsubsection{AI Enabled O-RAN Security}
With the introduction of xApp/rApp in Near-RT/Non-RT RICs and the establishment of data-driven closed-loop control within O-RAN, AI/ML is extensively employed to extract patterns from vast quantities of telemetry and signalling data, supporting anomaly detection and policy optimisation. However, the integration of AI/ML also expands the system's attack surface. Beyond traditional interfaces and cloud-native components, potential attack vectors include training data, feature engineering, model parameters, inference APIs, and the MLOps supply chain. O-RAN WG11's security updates \cite{oran_security_update_2025} indicate that O-RAN deployments have incorporated AI/ML-related threats into risk analysis, and further advanced the inclusion of model integrity, confidentiality, data poisoning protection, and other aspects in security requirements. Additionally, industry-led collaborative initiatives such as the AI-RAN Alliance have emerged, reflecting AI's long-term embedding within RAN control and optimisation processes. This development also implies that security design must be synchronously integrated into the AI lifecycle from the outset.

In the field of AI for Security, existing research primarily employs AI/ML as modules to enhance security situational awareness and detection capabilities. For instance, it is used for anomaly detection in O-RAN telemetry and sequence modelling of logs to identify abnormal calls and unauthorised access. However, prior studies have indicated that the openness of xApps/rApps within the O-RAN ecosystem may introduce novel adversarial vectors. For example, in scenarios with insufficient permission boundaries and isolation, malicious xApps may amplify their impact on RIC subsystems and network control by abusing interface calls, manipulating policy inputs, or inducing abnormal loads \cite{hung2024security}. Consequently, AI/ML security should be discussed together with foundational controls such as identity authentication, fine-grained authorisation, auditing, and compliance, rather than as a substitute for these fundamental security mechanisms \cite{ETSI_TS_104_104}.

In the field of Security for AI, O-RAN's AI/ML threats can reference general ML security risk classifications, such as the OWASP ML Top 10 \cite{OWASP_ML_Top10} and ENISA's categorisation of threats to ML systems \cite{baylon2021securing}. Integrating these with O-RAN's data closed-loop characteristics enables the development of more practice-oriented threat models. Typical risks include: training data poisoning and replay attacks (contaminating KPIs/logs to induce model bias), model poisoning (causing models to produce targeted misclassifications under specific triggers), adversarial examples (manipulating inputs during inference to induce erroneous decisions), model theft and member inference (leaking model assets and privacy information via inference interfaces or log side-channels), and AI supply chain attacks (malicious model packages, dependent components, or malicious xApp/rApp components).

In addition, in O-RAN 6G ecosystem, generative AI (GenAI) can be empowered across multiple nodes and functional layers. For instance, GenAI modules may generate policy recommendations and fault root cause reasoning within Non-RT RICs, or provide semantic-level network behaviour prediction and policy adjustments within near-RT RICs \cite{ORAN_GenAI_6G_2025}. Whilst these capabilities enhance network intelligence and automated operational efficiency, they also introduce novel security challenges. As GenAI model training and inference rely on extensive network logs, telemetry data, and policy histories, their inputs may encompass untrusted data sources. This risks prompt injection, semantic-level misdirection, or even misconfiguration of network policies. Unlike traditional discriminative machine learning, generative models lack explicit rule boundaries when outputting semantic text or policy recommendations in the control plane. This makes erroneous or malicious prompts, or context tampering, significantly harder to detect using conventional pattern-based detection techniques.

Existing work typically categorises AI security mitigation measures into three layers: data, model, and operational governance \cite{oran_security_update_2025}. At the data layer, emphasis is placed on trusted data sources and integrity verification, combined with anomaly detection and data cleansing to reduce the effectiveness of poisoning and replay attacks. At the model layer, focus lies on model integrity protection, access control, and rate limiting, supplemented by privacy-preserving inference or output constraints to mitigate member inference and model reverse engineering. At the operational governance layer, emphasis is placed on model version management, deployment approval and rollback mechanisms, gradual rollout, and continuous monitoring to control the scope of impact from model failures and enhance traceability. Among these, the governance and measurement approach proposed by NIST AI RMF \cite{NIST_AI_100_1} provides a reusable framework for integrating AI risks into organisational-level risk management, facilitating the expansion of AI security from isolated technical issues into auditable engineering processes.

Beyond security, AI/ML in O-RAN also raises issues of explainability and accountability. For security decisions, XAI research emphasises providing verifiable evidence (such as key feature contributions, anomaly trigger criteria, and decision confidence variations) while ensuring timeliness, thereby supporting operational personnel in conducting reviews and compliance audits. Furthermore, explanatory information should be integrated into log and evidence chain management to fulfil post-incident accountability and retrospective analysis requirements \cite{brik2024explainable}. In multi-vendor ecosystems, the chain of responsibility typically decomposes into: model/algorithm providers; xApp/rApp developers and deliverers; platform and runtime environment providers; and operational/integration entities bearing operational and approval responsibilities. Existing standards and risk frameworks favour anchoring accountability to auditable process evidence rather than relying solely on declarations of principle \cite{oran_security_update_2025}.

\begin{table*}[t]
\centering
\caption{Key AI/ML Threats and Common Mitigation Strategies in O-RAN}
\label{tab:aiml_threats_mitigation}

\small 
\renewcommand{\arraystretch}{1.35} 
\begin{tabularx}{\textwidth}{@{} p{2.6cm} >{\raggedright\arraybackslash}X >{\raggedright\arraybackslash}X >{\raggedright\arraybackslash}X @{}}
\toprule
\textbf{Threat Category} & 
\textbf{Primary Target / Scope} & 
\textbf{Typical Impact} & 
\textbf{Common Mitigation Strategies} \\
\midrule

\textbf{Data Poisoning / Replay} & 
KPI/Log/Feature pipelines; Training datasets. & 
Model bias; Long-term policy degradation. & 
Data lineage \& integrity checks; Data sanitization; Outlier isolation. \\
\addlinespace

\textbf{Model Poisoning / Backdoor} & 
Training phase parameters; Transfer learning models. & 
Targeted misclassification; Hidden backdoor triggers. & 
Model integrity verification; Robust training; Pre/Post-deployment comparison. \\
\addlinespace

\textbf{Adversarial Examples} & 
Inference inputs (Telemetry / Features). & 
Inference errors; Policy mis-triggering. & 
Adversarial training; Confidence gating; Input transformation \& filtering. \\
\addlinespace

\textbf{Model Stealing / Privacy Inference} & 
Inference APIs; Logs \& Side-channels. & 
Model IP theft; User/Data privacy leakage. & 
Access control \& rate limiting; Output constraining; Privacy-preserving inference. \\
\addlinespace

\textbf{AI Supply Chain Attacks} & 
Model packages/dependencies; xApp/rApp artifacts. & 
Persistent implantation; Lateral movement. & 
Signing \& dependency governance; Release auditing; Runtime isolation \& least privilege. \\
\addlinespace

\textbf{Drift \& Failure} & 
Long-running models. & 
Performance degradation; Unpredictable behavior. & 
Continuous monitoring; Drift detection; Canary rollout \& rollback strategies. \\

\bottomrule
\end{tabularx}
\end{table*}

\subsubsection{Zero-Trust Architecture Enabled O-RAN Security}

With the blurring of network boundaries and the widespread use of cloud services and IoT devices, traditional security models that rely on static, pre-defined rules and policies can be difficult to adapt to in the face of rapidly changing network environments and attack patterns. Zero Trust Architecture (ZTA) is based on the principle of ``never trust, always verify'', which requires strict authentication and authorization of any access request at any time, regardless of the part of the network from which the access request originates \cite{stafford2020zero}. The implementation of this model enables security policies to adapt more flexibly to dynamically changing network environments and threat landscapes, thus improving the security and resilience of the entire network. 

The adoption of the zero-trust principle is important for advanced network architectures such as O-RAN. Research by the O-RAN Alliance Working Group 11 (WG 11) has found that unauthorized API usage has become one of the major security threats to near real-time RICs (RAN Intelligent Controllers) \cite{oranSpec2024}. While WG 11 recommends the use of authentication and role-based access control (RBAC) to mitigate this risk, the credentials and tokens that authentication and authorization mechanisms rely on can be stolen, as has happened in high-profile organizations like Facebook and GitHub \cite{Jackson2022GitHubTokens}\cite{guan2019dangerneighbor}\cite{jiang2023oztrust}. Therefore, within O-RAN's open APIs and cloud-native runtimes, reliance solely on one-time authentication and static RBAC proves insufficient to mitigate credential compromise exploitation and lateral movement pathways. ZTA places greater emphasis on continuous reassessment during sessions, dynamic policy convergence, and workload-based micro-segmentation. Additionally lateral movement threats, a key tactic in the threat matrix summarized by Microsoft, also pose a serious challenge to O-RAN's near real-time RIC \cite{weizman2021secure}. Even if the initial point of intrusion is recognized, an attacker can continue to maintain access by hiding his or her whereabouts. This makes detecting lateral movement exceptionally difficult, and existing solutions, such as those relying on complex algorithms or machine learning (ML)-based approaches, often lead to huge computational overheads and are not applicable to near-real-time RIC environments with stringent real-time requirements. ZTA, on the other hand, as a solution to address security needs in networks with untrustworthy infrastructures, is well positioned to address the above threats and mitigate the risks associated with relying on information from untrustworthy vendors. Under the guidance of ZTA, every component, every data transmission and every service of O-RAN will be subject to strict security checks without trust, thus creating a more secure and reliable network environment for users and service providers.

The intelligent zero-trust architecture design proposed by Ramezanpour et al. \cite{Ramezanpour2022IntelligentO-RAN}\cite{Liyanage2023OpenOpportunities} provides enhanced security measures in untrustworthy network environments by integrating advanced artificial intelligence algorithms. This architectural design adopts a service-oriented philosophy and seamlessly integrates with the Open RAN architecture. In this architecture, three core components, Intelligent Gateway or Proxy (IGP), Intelligent Network Security State Analysis (INSSA), and Intelligent Policy Engine (IPE), work together to build a complete security defense mechanism. The IGP serves as a user artificial intelligence engine that analyzes network traffic through reinforcement learning and generates a security posture and preliminary risk assessment to provide users with environmental awareness. The application of federated learning allows the IGP to learn from data from different sources to build more comprehensive security models for its specific local environment, aiming to maintain a high level of subject confidence when accessing network resources. INSSA, on the other hand, analyzes the security status of the network through models such as graph neural networks and performs risk assessments for accessing resources in the network. INSSA utilizes anomaly detection mechanisms designed to identify potential attacks and provide continuous security protection for the system through real-time risk assessments. IPE uses artificial intelligence trust algorithms to make access authorization decisions based on subject privileges, security policy rules, network state, and access trustworthiness scores. IPE is expected to use reinforcement learning to maximize system availability with the principle of least privilege, make authorization decisions based on the combined assessment of IGP and INSSA through the Long Short-Term Memory (LSTM) network, and continuously monitor the security status of the session. The close collaboration of these three components provides a robust framework for implementing zero-trust security measures around the O-RAN architecture, optimizing security performance through an intelligent approach and providing new perspectives and approaches to zero-trust practices in O-RAN networks.

It should be emphasised that introducing federated learning and intelligent risk assessment into IGP/INSSA/IPE does not automatically eliminate the internal attack surface. Instead, it may transform the training and decision chain into a new primary target for attack. Firstly, federated learning on the IGP side may encounter data or model poisoning, backdoor updates, and Sybil participant manipulation during the aggregation phase. This could systematically distort risk assessment outcomes without triggering traditional alerts. Secondly, once policies or trust engines are compromised, attackers may execute malicious strategies under the guise of compliance, transforming policy execution points into risk amplifiers. Lastly, the cross-component state-sharing mechanism and trust-scoring data pathways may introduce collusion risks. For instance, when both the IGP and INSSA fall under joint control due to the same supply chain incident or runtime intrusion, attackers could coordinate to manipulate observation and assessment outcomes, constructing a superficially consistent low-risk state. This would mislead the IPE's authorisation and access decisions. Therefore, zero-trust discussions for O-RAN should not solely focus on external access controls but must also incorporate the training process, evidence chains, and component dependencies into threat modelling and audit boundaries.

Jiang et al. \cite{jiang2023oztrust} present OZTrust, an innovative zero-trust security system designed for O-RAN environments. OZTrust aims to address key security challenges in O-RAN Near-RT RICs, including unauthorized API access and lateral movement, and effectively protects sensitive and proprietary data from theft while preventing malicious manipulation of RAN controls. With two core components - the Access Control Module and the Policy Management Module - OZTrust minimizes system overhead while providing granular security protection. The access control module is built on the advanced eZTrust \cite{zaheer2019eztrust} technology, an eBPF-based access control mechanism designed for containerized environments. OZTrust has adapted and optimized it to fit the specific needs of O-RAN, in particular, to optimize the packet verification process by pre-populating the context mapping to avoid time-consuming processing paths. The Policy Management module is responsible for generating and enforcing access control policies for xApp. This process is vendor information independent and utilizes distributed trace libraries and tools to discover and validate xApp communication patterns to automate the export of access control policies. This approach not only reduces the risks that can be introduced by relying on third-party information, but also minimizes the potential for manual configuration errors.

The critical point to emphasise is that zero trust does not mean zero-cost security. Policy decision-making and execution pathways, session-level re-authorisation, and audit trails all introduce additional overhead and potential errors. This is important when operating under near-RT RIC constraints with strict latency and resource budgets, necessitating clear boundary definitions for trade-offs. Table \ref{tab:zta_metrics_near_rt} summarises typical overhead and trade-off dimensions.

\begin{table*}[t]
\centering
\caption{Key ZTA Metrics for Near-RT RIC Scenarios}
\label{tab:zta_metrics_near_rt}

\small 
\renewcommand{\arraystretch}{1.35} 

\begin{threeparttable}
    \begin{tabularx}{\textwidth}{@{} p{3.2cm} >{\raggedright\arraybackslash}X >{\raggedright\arraybackslash}X @{}}
    \toprule
    \textbf{Metric} & 
    \textbf{Definition (Focus on near-RT RIC context)} & 
    \textbf{Typical trade-off implication} \\
    \midrule

    \textbf{Policy Evaluation Latency} & 
    Additional latency introduced by the Policy Decision Point to complete context evaluation and decision-making (accounting for decision cache hits/misses). & 
    Richer context dimensions and constraints generally increase computational overhead; must be carefully balanced against the strict near-RT latency budget. \\
    \addlinespace

    \textbf{Per-request authorization overhead} & 
    Average overhead (CPU, memory, or instruction path) at the Policy Enforcement Point to execute authentication, authorization, and policy matching for a single request. & 
    Finer-grained access control typically incurs higher enforcement costs; requires optimization via hot-path analysis and caching strategies. \\
    \addlinespace

    \textbf{False Rejection Rate \& False Acceptance Rate} & 
    Proportion of legitimate requests rejected and illegitimate requests allowed, measured under fixed threat models and workloads. & 
    Stricter policies can reduce false acceptance but may raise false rejection; this should be co-designed with alerting, rollback, and human-in-the-loop review where necessary. \\
    \addlinespace

    \textbf{Policy Rollback Time} & 
    Time elapsed from triggering a rollback (after anomaly detection/misconfiguration) to the policy taking effect at the Policy Enforcement Point. & 
    Faster rollback reduces the blast radius, but depends on policy versioning, release gating, and automation in orchestration pipelines \\
    \addlinespace

    \textbf{Audit Coverage} & 
    Extent of traceability for critical decision and enforcement events (e.g., recording subject, object, action, context, and result). & 
    Higher audit coverage enhances forensics and accountability but significantly increases log collection, storage, and retrieval costs. \\

    \bottomrule
    \end{tabularx}
\end{threeparttable}
\end{table*}

\subsubsection{Quantum Communications and Post-Quantum Security in O-RAN}
Quantum communication technology utilizes the principles of quantum mechanics, such as quantum entanglement and quantum superposition, to achieve secure transmission of data, and it can be integrated with O-RAN to achieve high security and efficiency. For example, O-DUs and O-RUs can be obtained from different equipment vendors and require stringent timing requirements between them, so communications between them can be easily eavesdropped and can be desynchronized due to large delays. Quantum Key Distribution (QKD), on the other hand, serves as an unconditionally secure key exchange mechanism that allows two parties to securely share encryption keys in the presence of potential eavesdroppers \cite{Koashi2006UnconditionalPrinciple}. Using QKD, O-RAN components from different equipment vendors can be simply and securely connected, effectively preventing security threats such as eavesdropping and man-in-the-middle attacks \cite{Wang2022Quantum-EnabledChallenges}. In addition, some studies have demonstrated that optimal solutions can be achieved in wireless resource allocation by utilizing the quantum annealing algorithm on the D-Wave quantum computer, which greatly improves energy efficiency and spectrum efficiency \cite{Kim2019LeveragingNetworks}. QKD is currently being used to encrypt 5G-centric networks \cite{Wright20215GSecurity}, and China has already established a 4,600-kilometer QKD network for network traffic in 2021 \cite{pittaluga2021600}. Table \ref{tab:quantum_benefits} illustrates how several quantum technologies can benefit integration with O-RAN, highlighting challenges such as enhanced network security and efficiency, as well as implementation cost and complexity.

\begin{table*}[t]
\centering
\caption{Benefits of Quantum Technologies in O-RAN Integration}
\label{tab:quantum_benefits}
\begin{tabular}{>{\raggedright\arraybackslash}p{2cm} >{\raggedright\arraybackslash}p{3cm} >{\raggedright\arraybackslash}p{3cm} >{\raggedright\arraybackslash}p{3cm} >{\raggedright\arraybackslash}p{3cm}}
\hline
\textbf{Technology} & \textbf{Benefits to O-RAN Integration} & \textbf{Integration Level with O-RAN} & \textbf{Network Security Enhancement} & \textbf{Cost \& Complexity} \\
\hline
\textbf{QKD} \cite{Wang2022Quantum-EnabledChallenges} & Enhances data security across the network by providing a means for unconditionally secure key exchanges, vital for secure communications in heterogeneous networks. & Experimental and pilot projects underway. Limited commercial deployment primarily in secure government and financial networks. & Significantly increases due to unconditionally secure communication. Prevents eavesdropping and man-in-the-middle attacks. & High due to the need for specialized hardware and quantum repeaters for longer distances. Cost-effectiveness improves with advancements. \\ \hline
\textbf{Quantum Annealing} \cite{Kim2019LeveragingNetworks} & Optimizes network configurations and resource allocation more efficiently than classical algorithms, potentially reducing operational costs and energy consumption in network management. & Initial stages of integration. Used in R\&D for optimizing network configurations and resource allocation. & Indirectly enhances by optimizing network resource usage and efficiency, potentially reducing points of vulnerability. & High due to limited availability and complexity of integration. \\ \hline
\textbf{Quantum Sensors} \cite{bongs2023quantum} & Provides high-precision monitoring capabilities that could greatly enhance network diagnostics and fault detection, leading to more robust network operations. & Research phase with some early commercial applications outside telecommunications. Potential for network monitoring and optimization. & Could enhance network diagnostics and security by providing high-precision monitoring of physical and environmental conditions. & Initial costs are high, but decreasing as technology matures and production scales. \\ \hline
\end{tabular}
\end{table*}

Beyond QKD based on physical layer assumptions, another mitigation approach against quantum computing threats is post-quantum cryptography (PQC). Its core principle involves replacing key exchange and digital signature components within existing public-key systems that are vulnerable to Shor's algorithm \cite{11132566}, without relying on quantum channels or specialised optical hardware. This mitigates the `Harvest now, decrypt later' risk. NIST has published the lattice-based key encapsulation mechanism ML-KEM (FIPS 203) \cite{NIST_FIPS_203} and the lattice-based digital signature ML-DSA (FIPS 204) \cite{NIST_FIPS_204}, alongside the hash-based SLH-DSA (FIPS 205) as another category of quantum-resistant signature standards. In O-RAN scenarios, PQC is better suited for prioritised deployment on open interfaces and management control planes to achieve quantum-resistant reinforcement of critical control messages and policy distribution links. For high-throughput user plane data, reliance on symmetric cryptography remains primary, with PQC's impact primarily manifesting as handshake overhead and computational burden during session establishment. Regarding standardisation and interoperability, IETF is also advancing the use of ML-KEM for TLS 1.3 key agreement mechanisms \cite{IETF_Draft_TLS_MLKEM}, providing protocol-layer solutions for hybrid deployments. To further quantify the communication and computational costs of quantum-secure approaches in O-RAN deployments, Table \ref{tab:pqc_qkd_tradeoff} presents a comparative analysis of key overheads for traditional public-key systems, PQC, and QKD (primarily focusing on typical key sizes and deployment complexity during the session establishment phase).

\begin{table*}[t]
\centering
\caption{Quantified trade-offs of classical crypto, PQC, and QKD for O-RAN secure connectivity.}
\label{tab:pqc_qkd_tradeoff}

\small
\setlength{\tabcolsep}{4pt}
\renewcommand{\arraystretch}{1.25} 

\newcolumntype{L}{>{\raggedright\arraybackslash}X}
\begin{tabularx}{\linewidth}{@{} 
    >{\hsize=0.6\hsize}L 
    >{\hsize=0.9\hsize}L 
    >{\hsize=1.1\hsize}L 
    >{\hsize=1.4\hsize}L 
@{}}
\toprule
\textbf{Option} &
\textbf{Representative primitives} &
\textbf{Key/signature material (bytes)} &
\textbf{Deployment implications} \\
\midrule

Classical (baseline) &
TLS 1.3 key share (X25519) &
32\,B key share &
Mature ecosystem; minimal handshake overhead; vulnerable to large-scale quantum adversaries \\
\addlinespace

PQC (standardized) &
ML-KEM-768; ML-DSA-65 &
\textbf{ML-KEM-768:} 1184\,B encaps.\ key; 1088\,B ciphertext \newline
\textbf{ML-DSA-65:} 1952\,B public key; 3309\,B signature &
Software-upgradable on many nodes; increases handshake/cert payload (KB-level) and CPU cost; requires crypto-agility and interoperability testing \\
\addlinespace

QKD-based keying &
QKD for symmetric key distribution &
Not captured by message bytes (hardware-dependent); requires quantum channel and key management &
Strong information-theoretic properties but needs dedicated optical/QKD hardware, distance/rate constraints, and key management integration \\
\bottomrule
\end{tabularx}
\end{table*}

Although quantum communication and PQC can enhance the security of O-RAN/6G, it still faces multiple technical and implementation challenges in actual integration:
\begin{itemize}
    \item \textbf{QKD Scalability: }Existing solutions are mostly point-to-point \cite{Tsai2021QuantumSecurity}, and in the complex multi-point, multi-layer architecture of O-RAN, achieving multi-point key distribution and cross-layer synchronisation is extremely challenging; As network scale expands, it must also support large-scale key distribution and high-frequency updates for numerous endpoints; otherwise, practicality will be severely compromised.
    \item \textbf{Environmental Sensitivity: }Quantum devices require stringent conditions such as low temperatures and high stability. While kilometre-scale transmission has been achieved in recent years, and multi-mode quantum storage and quantum error correction have improved long-distance stability and efficiency, engineering deployment barriers remain extremely high \cite{liu20231002}\cite{lago2023long}.
    \item \textbf{Device Compatibility and Network Integration: }O-RAN architecture and quantum systems lack mature interface standards, resulting in high integration complexity and dedicated hardware costs \cite{Tsai2021QuantumSecurity}. Therefore, operators must carefully assess the true cost-effectiveness between security gains and deployment costs. In addition, PQC migration increases the key material and certificate signature load during the session establishment phase (at the kilobyte level). This places pressure on CPU/memory resources and handshake latency for resource-constrained edge nodes, requiring cryptographic agility, layered deployment, and interoperability testing.
\end{itemize}

\subsubsection{Digital Twin for Secure O-RAN}

Digital Twin (DT) technology has been increasingly applied in network systems. As virtual representations of physical entities, digital twins can precisely replicate the real-time operational states of these entities in real-world environments, enabling data collection, analysis, diagnosis, and prediction. The core of this technology lies in virtualizing real devices, networks, or systems; through real-time monitoring of their states and operations, it not only reflects current conditions but also analyzes historical and real-time data to predict possible future changes \cite{Jones2020Characterising}. In the context of O-RAN, DT is commonly referred to as DT-RAN. It leverages high-fidelity digital replicas of O-RAN elements, continuously synchronised with real-time telemetry data, to enable scenario analysis, secure testing of rApps/xApps, and closed-loop automation \cite{10179151}. Recent O-RAN research studies \cite{ORAN_nGRG_RS01_DT} further emphasise DT-RAN application scenarios such as Non-RT RIC/rApps testing, network automation, and site-specific optimisation, where the quality of security benefits largely depends on the fidelity and integrity of the data synchronisation loop.

\begin{figure}[h]
\centering
\includegraphics[width=3.5in]{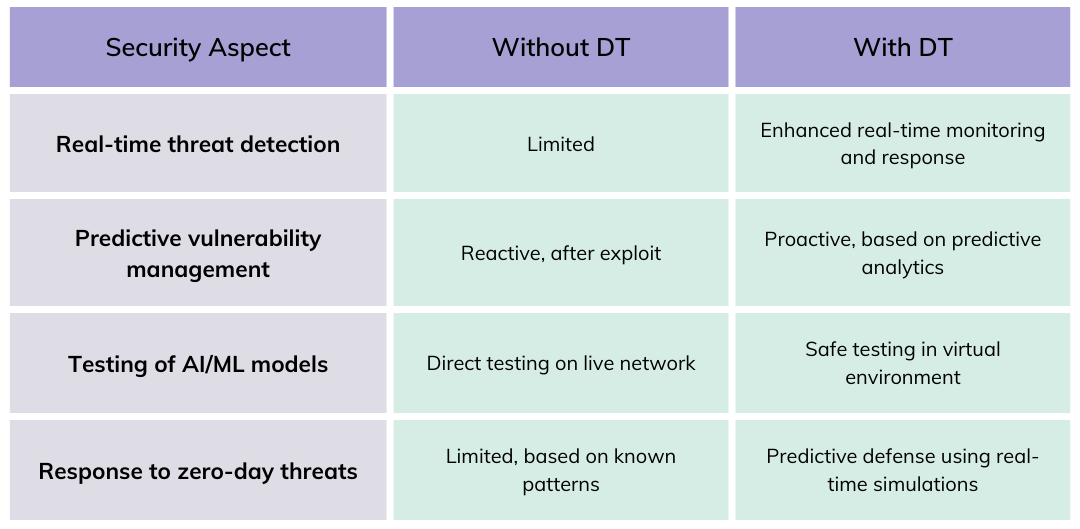}
\caption{Comparison of O-RAN security with and without Digital Twin technology}
\label{DT}
\end{figure}

As shown in Fig. \ref{DT}, the introduction of DT in O-RAN can enhance network security capabilities. On one hand, the DT environment can monitor network operational status in real time, proactively identify potential risks such as DDoS attacks and zero-day threats, and conduct predictive analysis and risk simulations through a simulation environment. On the other hand, new security policies and AI/ML models can be validated first in the DT virtual environment, avoiding the risks associated with direct deployment in the actual network. Additionally, DT can implement differentiated security management for network slices, dynamically adjusting encryption and access control policies in real time to enhance security isolation effects. Furthermore, by leveraging DT to compare the real-time status and historical data of network devices, network operations can transition from traditional post-failure repair to proactive early warning and self-healing, thereby improving the overall resilience and reliability of the network.

However, by introducing continuous data synchronisation loops and privileged analysis/simulation planes, DT also introduces new attack surfaces and implementation risks that should be modelled \cite{Kampourakis2025_DigitalTwin}:
\begin{itemize}
    \item \textbf{Twin drift:} Should synchronisation links be compromised (e.g., through data poisoning, replay attacks, or temporal distortion), the DT may diverge from the physical network state, leading to erroneous risk assessments or unsafe control decisions.
    \item \textbf{DT compromised as pivot point:} DT environments typically aggregate high-value assets (e.g., configurations, policies, models). Attackers may exploit this to steal configurations and sensitive data, tamper with or inject parameters, and propagate compromised policies back to the physical network, causing persistent impacts on real-time control and operational continuity.
    \item \textbf{Privacy exposure:} DT pipelines may ingest user-level operational data (e.g., location, slice-level KPIs). Without data minimisation and de-identification, DT may increase re-identification and inference risks, particularly in multi-tenant environments.
\end{itemize}

In recent years, research on digital twin technology in the O-RAN field has made significant progress, particularly in optimizing network management and enhancing network resilience. Current research primarily focuses on applying digital twins to key components of O-RAN systems, such as O-RUs, O-DUs, and Near-RT RICs, to create real-time virtual mirrors of these physical components \cite{mirzaei2023network}\cite{O-RAN_DT_Report_2024}. These studies emphasize that through real-time data interaction, digital twins can simulate and analyze various scenarios within the network, helping operators conduct risk assessments and performance optimizations without impacting the actual network.

\section{Standardization and Policy Development}
Ensuring a secure and reliable O-RAN deployment requires a strong standardization and policy framework to address system vulnerabilities and compliance requirements. International standards agencies and industry consortia provide common specifications that allow various components to interoperate efficiently while adhering to security best practices. By embedding security and privacy considerations into these standards, stakeholders can mitigate risk from the outset. Meanwhile, policy measures and regulations, such as data protection laws, guide O-RAN operators to protect user data and maintain trust.

\subsection{Key Standardization Organizations and Their Roles}
Multiple organizations drive O-RAN-related standardization, each with a specific role in security, privacy and compliance:
\begin{itemize}
    \item \textbf{O-RAN Alliance:} O-RAN Alliance \cite{oran2023security} is the primary consortium developing the O-RAN specification. Its Security Working Group (WG11) focuses on end-to-end security architecture and controls for O-RAN components and interfaces. For example, WG11 performs threat modelling and defines security requirements to ensure that open interfaces (A1, O1, E2, etc.) are protected by authentication, encryption and access control. The Alliance's Test and Integration Focus Group (TIFG) complements this by integrating a security testing framework and conducting interoperability testing. Through these efforts, O-RAN Alliance has laid the foundation for a secure multi-vendor RAN ecosystem. It also coordinates with other organizations through its Standards Development Focus Group (SDFG) to align O-RAN specifications with external standards.
    \item \textbf{3GPP:} 3GPP \cite{Cao2020ANetworks}\cite{Chen20235G-Advanced} is responsible for the development of broader mobile network standards, including a dedicated security group, SA3, which develops security specifications for 4G, 5G and evolving 6G networks. In the context of O-RAN (based on the 5G architecture defined by 3GPP), 3GPP SA3 defines measures such as mutual authentication of network elements, encryption of user data and signalling integrity protection. These standards cover security from the core network to the radio interface and ensure that basic mobile network protection measures (such as SIM authentication and air interface encryption) are implemented. 3GPP's work provides the basic security architecture within which O-RAN open interfaces must be protected, and it will continue to develop standards to address future threats and technologies.
    \item \textbf{ITU (International Telecommunication Union):} As a specialised agency of the United Nations, the ITU \cite{Jones1997The} promotes global telecommunications standards, including security and privacy. Within the ITU's Standardization Sector (ITU-T), Study Group 17 (SG17) works on standards for cybersecurity and privacy protection. ITU-T SG17 develops recommendations on topics such as cybersecurity management, secure authentication protocols and data protection guidelines. These global standards inform national regulators and industry participants to ensure consensus on security best practices. For O-RAN stakeholders, the ITU's work (e.g., Telecom Data Processing Guidelines or the Cybersecurity Framework) helps to ensure that O-RAN implementations comply with internationally recognised security and privacy principles. In essence, the ITU provides high-level security and privacy guidance to complement the technical specifications of 3GPP and the O-RAN Alliance.
\end{itemize}

O-RAN WG11 specifies security requirements for open interfaces and cloud-native RAN components, 3GPP SA3 defines the baseline security architecture for mobile systems (e.g., authentication, key hierarchy, user plane protection), while ITU-T SG17 provides overarching network security and privacy principles. For multi-vendor O-RAN deployments, the actual gap lies not in a lack of standards, but in the absence of a clear coordination perspective that maps interface-level controls to system-level security objectives and compliance expectations. Therefore, we have summarised a coordination matrix across Standards Developing Organizations (SDO), as shown in Table \ref{tab:cross_sdo_alignment}. This matrix highlights common control themes (identity, interface protection, platform hardening, supply chain assurance, auditability) and clarifies how O-RAN-specific controls inherit, extend, or implement 3GPP/ITU guidance.

\begin{table*}[t]
\centering
\caption{Cross-SDO alignment view for O-RAN security (illustrative).}
\label{tab:cross_sdo_alignment}
\small
\renewcommand{\arraystretch}{1.25}

\newcolumntype{Y}{>{\raggedright\arraybackslash}X}

\begin{tabularx}{\textwidth}{@{} l Y Y Y @{}}
\toprule
\textbf{Theme} & \textbf{O-RAN WG11 focus} & \textbf{3GPP SA3 focus} & \textbf{ITU-T SG17 focus} \\
\midrule

Identity \& access & 
Interface/service authN/authZ for RIC/SMO/O-Cloud & 
System-wide auth, key hierarchy, security architecture & 
Secure authentication principles, governance \\
\addlinespace

Interface protection & 
TLS/IPsec, mutual authentication, cert lifecycle & 
Signalling/user-plane protection, integrity/confidentiality & 
Security management guidance \\
\addlinespace

Cloud-native platform & 
O-Cloud hardening, isolation, runtime security & 
Security architecture assumptions for virtualized networks & 
Cybersecurity frameworks and best practices \\
\addlinespace

Supply chain assurance & 
SBOM, signing, vulnerability response evidence & 
Requirements impacting network function trust & 
Supply-chain and security management guidance \\
\addlinespace

Auditability \& evidence & 
Interface logs, policy/model provenance (RIC) & 
Security event visibility at system level & 
Accountability and compliance principles \\

\bottomrule
\end{tabularx}
\end{table*}

\subsection{Security Standardization in O-RAN}
Due to the disaggregated and open nature of the O-RAN architecture, it is critical to standardise security in O-RAN as this introduces new interfaces and potential attack surfaces. O-RAN Alliance has developed a comprehensive set of security specifications and protocols that specifically address O-RAN components and their interactions:
\begin{itemize}
    \item \textbf{O-RAN Security Protocols and Controls:} To protect communications between O-RAN open interfaces and infrastructure, the Alliance requires the use of authenticated security protocols \cite{oranSpec2024}. Interfaces such as the forward pass (between O-RUs and O-DUs) use protocols such as IPsec and Transport Layer Security (TLS) for encryption and integrity protection. In the most recent specification, O-RAN Alliance WG11 introduces support for TLS 1.2/1.3 and mutual authentication (using PKI and X.509 certificates) on the management and control interfaces. In addition, O-RAN nodes and applications use protocols such as SSH for management access and, where appropriate, Datagram TLS (DTLS). The certificate management framework supports the use of public key encryption and has adopted OAuth 2.0 for authorisation in service-based interactions. Together, these standards designed to provide confidentiality, integrity, and authentication for O-RAN architecture.
    \item \textbf{Interface and Platform Security:} Given the modular architecture of O-RAN, each interface and platform component (Near-RT RIC, Non-RT RIC, O-Cloud, O-DU/O-CU, etc.) must be hardened \cite{oranSpec2024}. The Alliance's security specification enumerates controls on key interfaces such as A1 (between Non-RT RIC and Near-RT RIC), E2 (Near-RT RIC to O-DU/O-CU) and O1/O2 (management interface). For example, any applications (xApp/rApp) connected to the RIC platform need to authenticate and authorise each other to prevent unauthorised control of RAN functions. Platform security is also taken into account: the O-Cloud (cloud platform hosting O-RAN functionality) security guidelines define segregation between tenants in virtualised environments as well as secure access to network functions. By standardising these measures, O-RAN ensures that all vendors implement security baselines (e.g., hardened containers, secure boot, runtime protection) across the ecosystem, thereby reducing weaknesses in multi-component deployments.
    \item \textbf{Software Supply Chain Security:} An emerging area of O-RAN standardization is the security of its software supply chain. The O-RAN network relies heavily on software from multiple sources (vendors, open source community), which raises concerns about software vulnerability and integrity. To address this issue, the O-RAN Alliance has introduced Software Bill of Materials (SBOM) requirements for all O-RAN software components \cite{oran_security_update_2025}\cite{Zahan2023Software}. Following guidance from the U.S. National Telecommunications and Information Administration (NTIA), each software release must list its open source and third-party components and their versions. This transparency enables operators to quickly identify whether known vulnerabilities (e.g., in open source libraries) could affect their O-RAN deployments. In addition, security requirements include vulnerability scanning and patch management procedures as part of the O-RAN software lifecycle. The Alliance is also developing a coordinated vulnerability disclosure programme to manage and remediate defects found in O-RAN products. These supply chain security initiatives, including SBOMs and security certification badges, are designed to increase confidence that O-RAN solutions are free from hidden malware or insecure dependencies.
\end{itemize}

Despite explicit protocol selection, a recurring security issue in multi-vendor O-RAN deployments is the `weakest-link' effect \cite{5137460}. This arises from security options dependent on specific implementations and inconsistent default settings. When different vendors configure encryption, certificate rotation, or authorisation scopes differently, the intended interface-level security assurance may be compromised by a single weak endpoint. While standards can provide clear specifications for the protocol stack, critical controls such as those mentioned above are often relegated to optional spaces. This lack of a unified mandatory baseline, coupled with the absence of requirements for testable and auditable compliance evidence, makes it difficult to systematically converge implementation divergences within multi-vendor environments.

\subsection{Data Privacy and Regional Compliance Requirements}
In addition to technical security standards, O-RAN systems must also comply with data privacy laws and regulations in different jurisdictions. Telecoms networks themselves process personal data (subscriber information, network usage data), so O-RAN operators need to ensure that their policies and designs comply with privacy regulations wherever their networks are deployed. Key regional frameworks include:
\begin{itemize}
    \item \textbf{General Data Protection Regulation (GDPR):} The EU's GDPR \cite{regulation2018general} is a comprehensive law governing personal data protection and privacy. The GDPR establishes principles such as data minimisation, purpose limitation and storage limitation, and enforces data subject rights, such as the right to access, correct and delete personal data. Any O-RAN deployment in the EU must incorporate these principles - for example, by ensuring that data collected by O-RAN rApps/xApps (which may analyse user device metrics or location information) is processed with user consent and fully anonymised or pseudonymised. The GDPR's stringent requirements for processing security (Art. 32) also mean that O-RAN operators in Europe must implement state-of-the-art security controls to protect personal data in transit and at rest. Non-compliance can result in significant fines, so O-RAN designs often include compliance checkpoints to enforce GDPR guidelines. 
    \item \textbf{American Data Privacy and Protection Act (ADPPA):} ADPPA \cite{uscongress2022hb8152} is a proposed piece of U.S. federal privacy legislation (as of 2022) that seeks to establish nationwide rules for the protection of personal data. While the U.S. currently lacks a law equivalent to the GDPR (relying instead on a patchwork of state laws and industry regulations), the ADPPA represents a trend toward stricter privacy oversight. Even in its proposed form, ADPPA indicates best practices (such as data subject rights and algorithmic accountability) that O-RAN operators may voluntarily adopt.
    \item \textbf{Other Regional Frameworks:} In addition to the EU and the U.S., other regions are introducing privacy laws that O-RANs must comply with. A notable example is India's Digital Personal Data Protection (DPDP) Bill (2023) \cite{dlapiper_dataprotection_india_2025}, which sets out requirements on how digital personal data should be processed, stored and transferred. With its emphasis on consent-based data processing and the imposition of penalties for data breaches, the DPDP Bill reflects similar principles to the GDPR, but is tailored to the Indian context. As a result, O-RAN networks operating in India will need to ensure that any personal data processed in the RAN or associated cloud (e.g. call detail logs) is securely stored in permitted areas and that users consent to the use of analytics or optimisation features. Similarly, countries such as Japan (APPI) \cite{dlapiper_dataprotection_japan_2025}, South Korea (PIPA) \cite{dlapiper_dataprotection_2025}, and Brazil (LGPD) \cite{dlapiper_dataprotection_brazil_2024} have their own privacy regulations that require compliance. The common theme is that the O-RAN architecture must be flexible enough to enforce region-specific data processing policies (e.g., enabling or disabling certain data collection features, or routing data to specific secure servers) to meet local law requirements.
\end{itemize}

\subsection{Ethical and Legal Challenges of AI Integration in O-RANs}
O-RAN's architecture makes full use of AI and ML, for example, to autonomously optimise the network in the form of rApps and xApps. While AI can improve efficiency and performance, it also poses significant ethical and legal challenges when deployed in telecoms environments:
\begin{itemize}
    \item \textbf{Algorithmic Transparency and Bias:} Many of the AI models used in O-RAN (for tasks such as traffic prediction, anomaly detection, etc.) operate as `black box,' especially complex deep learning models. This lack of interpretability leads to transparency concerns - stakeholders and regulators may question how decisions are made \cite{allahrakha2023ai} \cite{boza2021implementing} \cite{ebers2020regulating}. If AI-driven RAN functions cannot justify their actions, it is difficult to audit them for fairness or correctness. In addition, AI systems may inadvertently introduce algorithmic bias. If there are biases in the training data or model design, AI in a RAN may allocate unequal resources or degrade the quality of service for certain user groups, raising fairness and non-discrimination concerns \cite{Yampolskiy2020UnexplainabilityAI}. These concerns highlight the need for algorithmic transparency - for example, using explainable AI techniques or at least robust monitoring of AI decisions in the RAN. Standardization efforts are beginning to acknowledge this. For example, the O-RAN Alliance is exploring interpretability and bias mitigation requirements for AI/ML models used in O-RAN controllers.
    \item \textbf{Privacy Implications of AI:} The AI capabilities of O-RANs often require large data sets (e.g., detailed network telemetry, user device statistics) to learn and make decisions. This raises privacy concerns about how this data is handled. Even if personal identifiers are not used directly, the rich metadata analysed by AI may be sensitive information or reveal patterns of user behaviour. Ensuring that AI in O-RAN does not misuse personal data requires strict data governance and possibly privacy-preserving techniques (e.g., anonymisation or differential privacy). Ethical AI design in O-RAN requires assessing what data is really needed for AI tasks and minimising the exposure of user information.
    \item \textbf{Regulatory Gaps and Legal Uncertainty:} Currently, the deployment of AI in O-RAN is in a grey area with respect to telecom laws and AI-specific regulations. Traditional telecoms regulations (focusing on reliability, spectrum usage, etc.) do not foresee autonomous decision-making systems to manage network functions. At the same time, AI-specific regulations are only just emerging - for example, the EU's proposed AI Act \cite{eu_ai_act} would categorise and regulate high-risk AI systems, and it is conceivable that certain network management AIs could be categorised as high-risk if they have a significant impact on critical infrastructure or end-user services. However, as of 2025, the telecoms sector still lacks a clear legal framework for AI. Regulatory lag creates uncertainty for operators: liability issues (who should be responsible if AI-driven RAN optimisation leads to disruptions or violations of subscriber rights?), regulatory issues (what audits of AI algorithms might regulators require?). Without updated legislation, operators and providers will be exposed to legal risks or will be hesitant to fully embrace AI innovations \cite{allahrakha2023ai}. Similarly, data protection authorities may review how AI algorithms process user data under laws such as the GDPR's automated decision-making provisions. As a result, the lack of clear guidelines may slow the deployment of AI/ML in O-RAN. To address this issue, industry organisations and policymakers have begun drafting frameworks: the O-RAN Alliance liaises with regulators to share information on the use of AI in the network, and is conducting research into governance mechanisms for telecom AI.
\end{itemize}

\subsection{Practical Deployment Considerations in Multi-Vendor O-RAN}
The security posture of multi-vendor O-RAN systems depends on whether vendors adopt consistent security profiles for each interface and component. This includes:
\begin{itemize}
    \item Consistency in protocol and cipher suite selection (e.g., the use and version of TLS/mTLS)
    \item Cross-domain compatible PKI and certificate lifecycle management
    \item Interoperable authorisation semantics (e.g. OAuth2 token issuance)
    \item Version compatibility between different implementations (protocol stacks, API versions, and security libraries)
\end{itemize}
ETSI PAS provides a concrete benchmark for the transition to O-RAN WG11 specifications: Security Requirements and Control Specifications (TS 104 104) \cite{ETSI_TS_104_104}, Protocol Specifications (TS 104 107) \cite{ETSI_TS_104_107}, and Security Test Specifications (TS 104 105) \cite{ETSI_TS_104_105}. These specifications can be reused as acceptance criteria in interoperability testing, thereby reducing the likelihood of failure at the weakest link caused by inconsistent default values.

Under near-real-time constraints, a trade-off exists between performance and security. Although management and control interfaces should employ robust cryptographic protection, practical deployments need to budget for the overhead of mutual authentication handshakes, certificate validation, and session maintenance. TLS 1.3 reduces handshake round-trips and supports session recovery mechanisms, thereby lowering reconnection latency \cite{RFC8446}. This is crucial for enhancing operational resilience during component restarts or link jitter. O-RAN introduces near real-time control loops where unrestricted, per-message complex security processing may introduce operational risks. Effective solutions remain lightweight, establishing robust identity and security channels (e.g., mTLS) during session establishment, then avoiding placing costly cryptographic operations, complex authentication chains, or high-latency policy computations directly on millisecond-level control loops.

A CVE-2023-40998 \cite{TrendMicro_OpenRAN_Security} vulnerability exists in the Near-RT RIC implementation, allowing malformed messages from xApps to trigger a failure in E2 termination component, thereby causing a near-real-time control plane denial-of-service. CVE-2023-40997 \cite{TrendMicro_OpenRAN_Security} describes a maliciously crafted packet that causes E2 termination component to crash during decoding and memory operations, thereby disrupting subscription processing and degrading RIC availability. During multi-vendor integration, near-real-time RICs and E2 endpoints from different manufacturers are deployed with partially inconsistent security profiles. Although the design assumes protected transmission and strict input validation for E2 interface, operational pressures may lead to inconsistent protection mechanisms enabled across endpoints to minimise control loop overhead.

In practice, security variations across multi-vendor O-RAN deployments frequently manifest in observable outcomes such as authentication/authorisation overhead, certificate lifecycle management, configuration consistency, and auditability. To facilitate repeatable and comparable security assessments within multi-vendor O-RAN environments, we have summarised in Table \ref{tab:oran_deployment_kpis} a set of key metrics related to interoperability, performance overhead, and operational verifiability.

\begin{table*}[t]
\centering
\caption{Metrics candidates for comparable evaluation in multi-vendor O-RAN deployments.}
\label{tab:oran_deployment_kpis}

\small
\renewcommand{\arraystretch}{1.25}

\newcolumntype{Y}{>{\raggedright\arraybackslash}X}

\begin{tabularx}{\textwidth}{@{} Y Y Y Y @{}}
\toprule
\textbf{KPI} & \textbf{Definition} & \textbf{Why it matters} & \textbf{Typical measurement points} \\
\midrule

Policy evaluation latency &
Additional latency introduced by PDP/PEP decision and enforcement &
Captures the security gatekeeping cost under near-RT constraints &
RIC policy engine / PEP logs; control-plane timestamps (e.g., E2/A1) \\
\addlinespace

Mutual authentication handshake overhead &
Session setup/reconnect cost for mTLS/TLS/IKE (incl.\ resumption effectiveness) &
Affects scale-out, healing, and rolling upgrades stability &
TLS/IKE statistics; connection setup time; reconnect frequency \\
\addlinespace

Certificate rotation success rate / mean time to rotate &
Success ratio and time-to-complete for certificate renewal/rollover &
Reflects PKI lifecycle robustness and cross-vendor alignment maturity &
PKI/cert manager logs; component health signals and alerts \\
\addlinespace

Security configuration drift count &
Number of deviations from the agreed security profile/baseline over time &
Direct proxy for weakest-link risk in multi-vendor environments &
IaC/policy repo diffs; K8s admission/OPA; config audit reports \\
\addlinespace

Audit log completeness rate &
Coverage of required security events/fields for forensics and compliance &
Determines incident triage speed and accountability evidence quality &
SIEM/log pipeline; audit-field coverage checks \\

\bottomrule
\end{tabularx}
\end{table*}

\section{Conclusion}
The exploration and implementation of O-RAN technology represents a key evolution in mobile network architecture, providing unprecedented flexibility, intelligence and interoperability. By decoupling hardware and software, facilitating multi-vendor environments, and integrating advanced technologies such as artificial intelligence and machine learning, O-RAN facilitates a more competitive, innovative, and resilient telecommunications ecosystem. In addition, the transition to open and interoperable networks is expected to accelerate with the development of 6G, which promises to improve network performance, efficiency, and scalability.

However, the adoption of O-RAN also brings new challenges, especially in terms of security and privacy. The openness of O-RAN may increase the risk of cyber threats, and strong, standardized security measures and continued collaborative efforts by the global research and industry communities are needed to mitigate these risks. We provide a comprehensive overview of the different security threats and privacy challenges facing O-RANs and discuss in detail the security benefits of O-RANs and how they can be defended against security threats.

In conclusion, while O-RAN offers substantial benefits such as reduced costs, enhanced service delivery and improved user experience, its full potential can only be realized through careful management of the associated risks and challenges. The success of O-RAN in the 6G era will depend on the industry's ability to maintain high security standards and foster a regulatory environment that supports technological innovation and protects against emerging cyber threats. Going forward, the convergence of O-RAN and 6G technologies will unlock new possibilities for smart applications and services, further driving global digital transformation.

\bibliographystyle{IEEEtran}
\bibliography{references}

\begin{IEEEbiography}[{\includegraphics[width=1in,height=1.25in,clip,keepaspectratio]{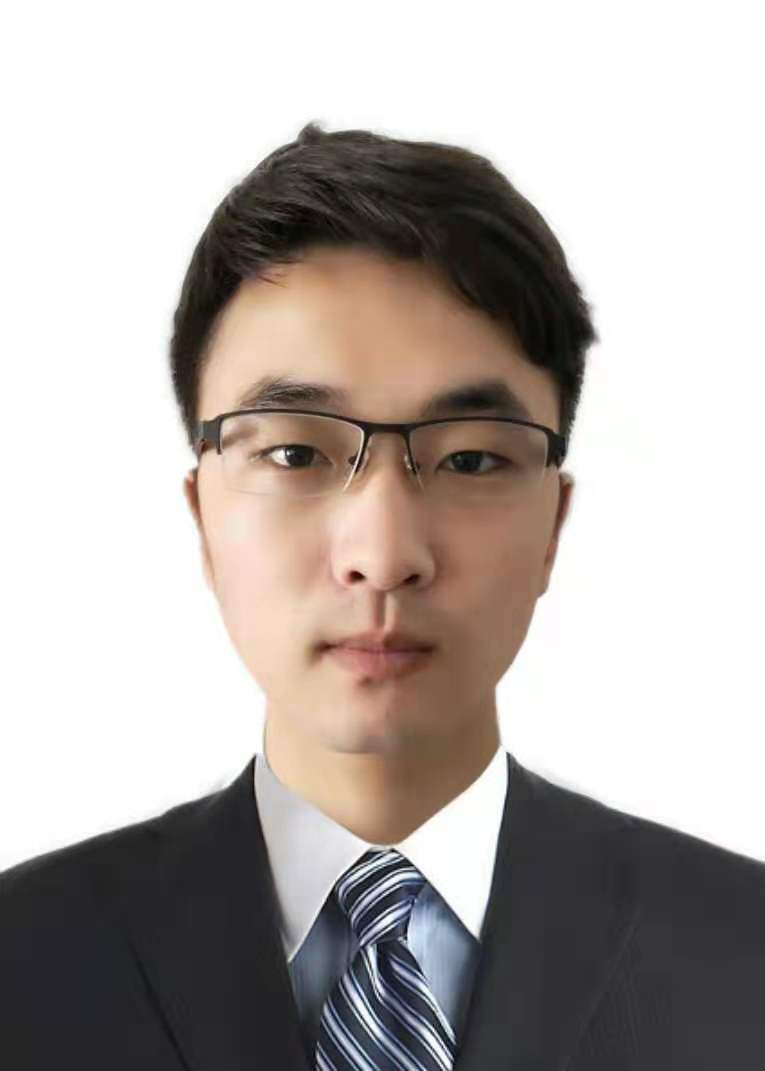}}]{Lujia Liang}
received the B.S. degree in information security from China University of Mining and Technology, Xuzhou, China, in 2021, and the M.S. degree in cybersecurity, privacy and trust from The University of Edinburgh, Edinburgh, U.K., in 2022. He is currently pursuing the Ph.D. degree with the James Watt School of Engineering, University of Glasgow, Glasgow, U.K. His research interests include network security, privacy and 6G networks.
\end{IEEEbiography}

\begin{IEEEbiography}[{\includegraphics[width=1in,height=1.25in,clip,keepaspectratio]{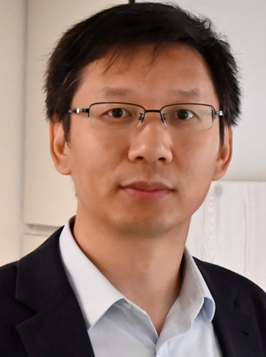}}]{Lei Zhang}
(Senior Member, IEEE) is a Professor of Trustworthy Systems at the University of Glasgow. He has academia and industry combined research experience on wireless communications and networks, and distributed and trustworthy systems for IoT, blockchain, and autonomous systems. His 20 patents are granted/filed in 30+ countries/regions. He published 3 books, and 150+ papers in peer-reviewed journals, conferences and edited books. Prof. Zhang is an associate editor of IEEE Transactions on Network Science and Engineering, IEEE IoT Journal, IEEE Wireless Communications Letters and Digital Communications and Networks, and a guest editor of IEEE JSAC, and IEEE Network Magazine. He received the IEEE Internet of Things Journal Best Paper Award 2022, IEEE ComSoc TAOS Technical Committee Best Paper Award 2019 and IEEE ICEICT’21 Best Paper Award. Dr. Zhang is the founding Chair of IEEE Special Interest Group on Wireless Blockchain Networks in IEEE Cognitive Networks Technical Committee (TCCN). He delivered tutorials in IEEE ICC'20, IEEE PIMRC'20, IEEE Globecom'21, Globecom'22, IEEE VTC'21 Fall, IEEE ICBC'21 and EUSIPCO'21.
\end{IEEEbiography}

\vfill

\end{document}